\documentclass[amstex,aps,amsfonts,prsty,preprint]{article} 
\usepackage[dvips]{graphicx}
\usepackage{amssymb}
\usepackage{amsmath}
\usepackage{a4wide}
\usepackage{citesort}

\def\be{\begin{equation}}
\def\ee{\end{equation}}

\begin{document}

\begin{titlepage}

\begin{center}

{\Huge \bf  Directed random walk in adsorbed monolayer
}

\vspace{0.3in}

{\Large \bf O.B{\'e}nichou$^{1}$, A.M.Cazabat$^{2}$, 
M.Moreau$^{1}$
 and G.Oshanin$^{1}$
}

\vspace{0.1in}

{\large \sl $^{1}$ Laboratoire de Physique Th{\'e}orique des Liquides (CNRS - UMR 7600), 
Universit{\'e} Pierre et Marie Curie,
4 place Jussieu, 75252 Paris Cedex 05, France\\
}
\vspace{0.05in}
{\large \sl $^{2}$ Laboratoire de Physique de la Mati{\`e}re Condens{\'e}e,
Coll{\`e}ge de France, 11 place M.Berthelot, 75252 Paris Cedex 05, France
} 
                 
\vspace{0.1in}

\begin{abstract}

We study the dynamics of a tracer particle, which performs
a totally directed random walk in 
an adsorbed monolayer 
 composed of mobile
hard-core particles undergoing
 continuous exchanges 
with a vapour phase. In terms of
  a mean-field-type
approach, based on the decoupling 
of the tracer-particle-particle 
correlation functions 
into the product of 
pairwise, tracer-particle correlations,  
we determine the
density profiles of the monolayer particles, 
as seen from the stationary moving tracer,  and
calculate
its terminal velocity, $V_{tr}$.  
In the  general case the latter is determined implicitly, 
as the solution of a certain transcendental equation.
 In two extreme limits of slow and fast monolayer
 particles diffusion, we obtain explicit asymptotic
 forms of $V_{tr}$. We show next that
 the density profile in the monolayer is
strongly inhomogeneous:
In front of the stationary
 moving tracer the local density 
is higher than the average value, $\rho_L$, and approaches $\rho_L$
as an exponential function of the distance from the tracer.
Past the tracer the local density is lower than $\rho_L$ and
 the approach to $\rho_L$ may proceed differently depending whether the particles
number in the monolayer is not or is explicitly conserved. In the former case
the approach is described  by  an exponential
 dependence 
with a different characteristic length, compared to the behavior in front of the tracer;
  in the latter case, the density tends to $\rho_L$ algebraically. 
 The characteristic lengths
 and the amplitudes of the density relaxation functions
 are also determined explicitly.

\end{abstract}

\vspace{0.1in}

\vspace{0.1in}
\end{center}

\end{titlepage}

\pagebreak

\section{Introduction}

When a gas  or vapour is brought into contact
with a clean solid surface, parts of its molecules 
will become reversibly attached to the surface in the form of an adsorbed layer. 
Knowledge of equilibrium and dynamical properties of such
adsorbed layers is essential for many practical applications,
including coating, gluing and lubrication.

Following the seminal
work of Langmuir (see, e.g. Ref.\cite{surf}), the
equilibrium
 properties of  adsorbed
layers 
have been extensively studied 
and a great number of
important developments have been made. 
In particular, 
further analysis
 included more realistic forms of  
intermolecular interactions 
or allowed for the possibility of
multilayer formation.  As a result, 
different possible phase transformations
have been predicted and 
different forms of
adsorption isotherms 
have been found,  which well explain
available experimental data 
(see, e.g.  Refs.\cite{surf,shaw,fowler}).  

Some effort has been also invested to the 
understanding of molecular diffusion 
in adsorbed layers,  which has a strong impact on
their global dynamical behavior. 
For instance, diffusional processes 
control the rates of spreading of molecular 
films on solid surfaces \cite{spreading1,spreading}, 
spontaneous or forced dewetting of monolayers 
\cite{aussere,persson,dewetting,dewetting2}
or island formation \cite{islands}.  
Here, some approximate analytical
results have been obtained for
both dynamics of an isolated adatom
on a corrugated surface and collective
diffusion, describing spreading of the macroscopic 
density fluctuations in
interacting adsorbates being in contact with 
the vapour phase
\cite{gomer,kreuzer,gortel,wahn}. 
On the other
hand, most of
available studies of the
tracer diffusion  
in adsorbed layers, which provides a useful information about
 their intrinsic viscosity,
pertain to  strictly two-dimensional models
 excluding the possibility 
of particles adsorption and desorption 
(see, e.g. 
Refs.\cite{kehr,elliott,tahir,beijeren,hilhorst,deconinck}
and references therein). 
Except for Ref.\cite{olivier}, which studied driven 
tracer dynamics in a somewhat artificial
one-dimensional model, analysis of the tracer diffusion 
in adsorbed monolayers undergoing 
exchanges with the vapour phase
seems to be lacking at present.

In this paper we study  the 
diffusion properties of a driven tracer 
in a two-dimensional adsorbed monolayer undergoing continuous exchanges
with the vapour. The system we consider 
consists of (i)
a solid substrate, which is modeled
in a usual fashion as a regular
square lattice of adsorption sites 
supporting at most a single occupancy, (ii) 
a monolayer of adsorbed, mobile
hard-core particles in contact with a vapour 
and (iii) a single, hard-core tracer
particle. Each  monolayer particle moves randomly
along the lattice
by performing 
hopping motion
on the vacant neighboring lattice sites.
The monolayer particles undergo continuous exchanges with a vapour phase,
i.e. may desorb from and adsorb onto the lattice
with  some prescribed rates dependent on the vapour 
pressure, temperature
 and the interactions with the solid substrate.
On the contrary, the
tracer particle is constrained to move 
on the two-dimensional
lattice only, 
i.e. it can not desorb 
to the vapour, and it is
subject to
a constant external force $E$. Hence, 
the tracer performs
a biased random walk,  
which is
constrained by the  hard-core
interactions with the
monolayer particles, 
and always remains within 
the monolayer, probing 
its response on the internal perturbancy or, in other words,
its frictional properties. 
Here we focus on the limit of 
sufficiently large external
force, such that the 
tracer may move only in one direction, but not large enough to make it 
slide regardless of the surface corrugation. 
The results for the general case of arbitrary $E$,
 which allow us to deduce the analog of the
Stokes formula for 2D adsorbed monolayers and to
 define the corresponding friction coefficient,
will be presented elsewhere \cite{ol}.

In terms of a mean-field-type
approximation of Ref.\cite{burlatsky}, which is based on
the decoupling of 
the tracer-particle-particle
 correlation functions into the product of pair-wise 
correlations, 
we define the
stationary density profiles of the monolayer particles, 
as seen from the moving tracer, and we
determine
analytically
the terminal velocity $V_{tr}$ of the tracer. We show that the 
adsorbed monolayer particles
tend to accumulate in front of the driven tracer, 
creating a sort of a "traffic jam", which impedes
 its motion. Thus the density profile  around the tracer is
highly inhomogeneous: the local density of the monolayer
 particles in front of the tracer is higher than the 
average and approaches the average value as an exponential
 function of the distance from the tracer. The characteristic
length and the amplitude of the density relaxation function
are calculated explicitly.
On the other hand, past the tracer 
the local density is lower than the average; we show that depending on
 the condition whether the number of particles in the monolayer 
is explicitly conserved or not, the local density past the tracer
 may tend to the average value either as an exponential or  as an
 $\it algebraic$ function of the distance, revealing in the latter case
especially strong memory effects and strong 
correlations between the particle distribution in the
 monolayer and the tracer position. 

Further on, we find
 that the terminal velocity of the tracer particle depends
 explicitly on the excess density in the "jammed" region in
 front of the tracer. This excess density, in turn,  depends on the
 magnitude of the velocity, 
as well as on the rate of the adsorption/desorption processes and on the
 rate at which the particles can diffuse away of the tracer.
 The interplay between the jamming of the monolayer, produced by the tracer 
particle, and the rate of its
homogenization due to diffusion and adsorption/desorption processes, manifests itself
 as a medium-induced frictional force exerted on the tracer, whose 
magnitude depends on the tracer velocity. 
As a consequence of such a non-linear coupling,    
in the general case, (i.e. for arbitrary adsorption/desorption 
rates and particles diffusion coefficient), 
$V_{tr}$ can be found only implicitly, 
as the solution of a certain non-linear 
equation relating $V_{tr}$ to the system parameters. 
This equation 
simplifies considerably   
in the limit of small or large 
particles diffusivity, in which two cases
exlicit asymptotic expressions for 
the tracer velocity are obtained. We finally remark,
 that a qualitatively similar physical effect was predicted recently
 for a  different model system involving a charged particle moving 
at a constant speed at a small distance above the surface of an incompressible,
 infinitely deep liquid. It has been shown in Refs.\cite{elie1,elie2}, 
that the interactions between the moving particle and the fluid molecules
  induce an effective frictional force exerted on the particle, producing
 a local distortion of the liquid interface, - a bump, which travels 
together with the particle and increases effectively its mass.  The mass of the bump, 
which is
analogous to the jammed region appearing in our model, depends itself
on the particle's velocity resulting
in a non-linear coupling  between the medium-induced
 frictional force exerted on the particle and its
velocity \cite{elie1,elie2}.

The paper is structured as follows: In Section
 2, we formulate
the model and introduce
basic notations.  In Section 3, we write down the dynamical
equations which govern the time evolution of the monolayer
particles and of the tracer. Section 4 is devoted to the
analytical solution of these
evolution equations in the limit $t \to \infty$ ; here we also
 present some 
general results on the form of the density profiles around stationary moving tracer and  
on the tracer terminal velocity,
which is given implicitly, as the solution of a
transcendental equation. In Section 5,
we derive explicit asymptotic results for the tracer terminal velocity 
in the limits of small and large particles diffusivities. As well, here we discuss the
forms of the density profiles at intermediate scales.
 Corresponding asymptotical behavior of
the density profiles at large distances from the tracer 
is discussed in Section 6. 
Finally, we conclude in Section 7 with a brief summary
and discussion of our results.

\section{The model}

Consider a
two-dimensional  square
 lattice of adsorption sites $(X,Y)$, 
which is brought 
in
contact with a reservoir of identic,
electrically neutral particles (vapour phase) (Fig.1), 
maintained at a constant pressure. 
We suppose that the particles may leave
 the reservoir and adsorb 
onto any vacant lattice site at a fixed rate $f/\tau^*$. Then, 
the adsorbed particles may move randomly on the lattice by  
hopping with a rate $l/ 4 \tau^*$ to any of four
neighboring lattice sites,
which process is 
constrained by hard-core exclusion preventing multiple occupancy of
any site. They can also 
 desorb from the lattice  back to the reservoir
at a site- and environment-independent rate $g/\tau^*$.
We assume, for simplicity of exposition, 
that typical adsorption, desorption and jump times
are equal to each other\footnote{These times, $\tau_{ad}$,
$\tau_{des}$ and 
$\tau_{jump}$, respectively, can be readily made different and
restored in our final results by the mere replacement
 $f \to \tau^{*} f/\tau_{ad }$, $g \to \tau^{*} g/\tau_{des}$ and
$l \to l \tau^*/\tau_{jump}$.}  
and are denoted by $\tau^{*}$. Consequently, at any time instant 
the 
variable $\eta(X,Y)$, characterizing the occupation of the site $(X,Y)$ such that
 
\begin{equation}
\eta(X,Y) = \left\{\begin{array}{ll}
1,     \mbox{ if the site $(X,Y)$ is occupied by an adsorbed particle} \nonumber\\
0,     \mbox{ if the site $(X,Y)$  is empty,}
\end{array}
\right.
\end{equation} 
can change its value due to adsorption,
 desorption and constrained random hopping events. 
Note
that the total number of 
particles in the adsorbed monolayer 
is not conserved in this dynamical
process. However, the mean density
 $\rho_L$ of the adsorbate, $\rho_L = <\eta(X,Y)>$, approaches as
$t\to\infty$ a constant value

\begin{equation}
\rho_L=\frac{f}{f+g},
\label{rhostat}
\end{equation} 
which equation is usually referred to as the Langmuir adsorption
isotherm \cite{surf}.

Further on, at $t = 0$ we introduce at the lattice origin 
an extra hard-core 
particle, whose motion
 we would like to follow and 
which will be called in the remainder as the tracer.
We stipulate that the tracer is different from the adsorbed particles
in that it can not desorb from the lattice and in that it is subject
to some external driving force, which favors its jumps to a
preferential direction. Such a  situation may be realized, 
for instance,
if this only particle is charged and the system is subject to a
uniform electric field. Here we will focus on the particular case
when the external field, oriented in the positive $X$-direction,  
is sufficiently strong, such that the tracer
performs totally directed random walk (Poisson process) along the $X$-axis 
in the two-dimensional monolayer of adsorbed
 mobile,
hard-core particles. 
 The tracer can be 
thought off as a certain 
external probe, which is designed to
measure the resistance offered by the monolayer
 particles to the external perturbance.

More precisely, we define the tracer  dynamics as
follows:
We suppose
that the tracer, which is at the site $X_{tr}(t)$ at time $t$,
 waits an exponentially distributed time with mean\footnote{We suppose that in the general case
this mean time  $\tau$ is different from the corresponding time $\tau^*$ 
associated with the monolayer particles dynamics. As a matter of fact,  this should be the case merely because the
tracer-substrate interactions may be different from the particle-substrate ones. Varying
 $\tau$ we can mimic different possible situations; in particular, $\tau= 0$ corresponds to the
case when the tracer simply slides along the substrate
 regardless of the surface corrugation. } $\tau$, 
and then attempts to hop onto the neighboring site $X_{tr}(t) + \sigma$,  $\sigma$ 
being the lattice spacing.
 The jump is actually fulfilled only if 
the target site
is vacant at this moment of time; otherwise, i.e., if the target site
is occupied by any adsorbed particle, the jump is rejected and the
tracer remains at its position.

\section{Evolution equations}

Let $P(X_{tr},\eta;t)$ denote the probability of finding  
at time $t$ the tracer at the site $(X_{tr},0)$
and all adsorbed particles in the configuration $\eta =
\{\eta(X,Y)\}$.
Further more, let $\eta^{(X,Y);(X',Y')}$ denote the
configuration obtained from $\eta$ by exchanging the occupation
variables of the two neighboring 
sites $(X,Y)$ and $(X',Y')$, which describes the Kawasaki-type
particle-vacancy exchange between the sites $(X,Y)$ and $(X',Y')$.
Next, we denote as 
$\eta^{(X,Y)}$ the configuration obtained from 
the original configuration $\eta$ by
the replacement $\eta(X,Y) \to 1-\eta(X,Y)$, which corresponds to the
Glauber-type flip of the occupation variable due to 
the adsorption/desorption 
events. 

Then, counting up all events which can result in the configuration 
$(X_{tr},\eta)$ at time $t$ or modify it, we write down the following
master equation, which governs the time evolution of the configuration
probability $P(X_{tr},\eta;t)$:

\begin{eqnarray}
\dot{P}(X_{tr},\eta;t)&=&\frac{l}{4\tau^*}\sum_{
(X,Y)}  {^\prime} {^\prime} \; 
\{P(X_{tr},\eta^{(X,Y);
(X+\sigma,Y)};t)-P(X_{tr},\eta;t)\} +\nonumber\\
&+&\frac{l}{4\tau^*}\sum_{
(X,Y)}  {^\prime} {^\prime} \; 
\{P(X_{tr},\eta^{(X,Y);
(X,Y+\sigma)};t)-P(X_{tr},\eta;t)\} +\nonumber\\
&+&\frac{1}{\tau}\{(1-\eta(X_{tr},0))P(X_{tr}-\sigma,\eta;t)-
(1-\eta(X_{tr}+\sigma,0))P(X_{tr},\eta;t)\} +\nonumber\\
&+&\frac{g}{\tau^*}\sum_{(X,Y)}  {^\prime}  \;
\{(1-\eta(X,Y))P(X_{tr},\eta^{(X,Y)};t)
-\eta(X,Y)P(X_{tr},\eta;t)\} +\nonumber\\
&+&\frac{f}{\tau^*}\sum_{(X,Y)}  {^\prime} \;
\{\eta(X,Y)P(X_{tr},\eta^{(X,Y)};t)
-(1-\eta(X,Y))P(X_{tr},\eta;t)\},\nonumber\\
\label{eq maitresse}
\end{eqnarray} 
where the dot denotes the time derivative, double prime indicates that the summation extends over all lattice sites
$(X,Y)$, excluding the special sites $(X_{tr}-\sigma,0)$ and 
$(X_{tr},0)$; a single prime after the summation
sign means that the sums runs over all lattice sites
excluding the site $(X_{tr},0)$ only.

Now, we can obtain the instantaneous 
velocity $V_{tr}(t)$ 
of the tracer particle directly from Eq.(\ref{eq maitresse}).
To do this, we multiply both sides of Eq.(\ref{eq maitresse}) by
$X_{tr}$ and sum over all possible configurations $(X_{tr},\eta)$, 
which yields the following equation for the instantaneous tracer velocity
\begin{equation}
V_{tr}(t)=\frac{d\overline{X_{tr}(t)}}{dt}=
\frac{\sigma}{\tau}(1-k(\sigma,0;t)), 
\label{vitesse}
\end{equation}
where the bar denotes the configurational average and
\begin{equation}
k(\lambda_1,\lambda_2;t)=\sum_{(X_{tr},\eta)}\eta(X_{tr}
+\lambda_1,\lambda_2)P(X_{tr},\eta;t)
\label{def de k}
\end{equation}
is the probability of having at time t an adsorbed particle 
at position $(\lambda_1,\lambda_2)$, defined 
in the frame of reference moving with the
tracer particle. In other words, $k(\lambda_1,\lambda_2)$
 can be interpreted 
as the density
profile as seen from the moving tracer. 

Consequently, Eq.(\ref{vitesse}) relates 
the instantaneous velocity of the tracer
 particle to the particle density in the 
immediate vicinity of the tracer. Note, 
that if the monolayer was perfectly stirred, i.e. 
$k(\lambda_1,\lambda_2) = \rho_L$ everywhere, (which implies 
 decoupling of $X_{tr}$ and $\eta$),  we should obtain
from Eq.(\ref{vitesse})  a trivial mean-field result

\begin{equation}
\label{mean-field}
V_{tr} = \frac{\sigma   (1 - \rho_L)}{\tau},
\end{equation}
which states that the tracer jump time $\tau$ is merely renormalized by a
 factor $(1 - \rho_L)^{-1}$, which is the inverse 
concentration of voids in the monolayer:  $(1 - \rho_L)/\tau $ defines simply 
the frequency of successful jump events. 
However, this
 turns out not to be the case and $k(\lambda_1,\lambda_2)$ appears to be different of the equilibrium value $\rho_L$ everywhere, except for infinitely separated sites, $\lambda_{1,2} \to \pm \infty$. Moreover, as we proceed to show, $k(\lambda_1,\lambda_2)$ is
itself dependent 
on the tracer velocity which yields ultimately  a 
strongly non-linear equation determining $V_{tr}$.

Hence, in order to calculate the mean
velocity of the tracer we have to determine the form of
the density profile $k(\lambda_1,\lambda_2)$, or more precisely, 
the value of the mean density 
at the neighboring to the tracer site
$(\lambda_1=\sigma,\lambda_2=0)$. The latter can be found
 from the master equation     
(\ref{eq maitresse}) by multiplying both sides of Eq.(\ref{eq maitresse})
by $\eta(X_{tr} + \lambda_1; \lambda_2)$ and performing the summation over all
configurations. 

In doing so, we
find the following set of equations, which hold for the sites
 separated from the tracer by the
distance exceeding  the lattice spacing $\sigma$, i.e. such that
$\lambda_{1}^2 +
\lambda_{2}^2 > \sigma^2$, 

\begin{eqnarray}
\dot{k}(\lambda_1,\lambda_2;t)=
\frac{l}{4\tau^*}\{\triangle_{\lambda_1}+\triangle_{\lambda_2}\}
k(\lambda_1,\lambda_2;t) -\frac{f+g}{\tau^*}k(\lambda_1,\lambda_2;t)
+\frac{f}{\tau^*}+\nonumber\\
+\frac{1}{\tau}\sum_{(X_{tr},\eta)}(1-\eta(X_{tr}+\sigma,0))
P(X_{tr},\eta;t)\{\eta(X_{tr}+\lambda_1+\sigma,\lambda_2)-
\eta(X_{tr}+\lambda_1,\lambda_2)
\},
\label{lambda>sigma}
\end{eqnarray}
where the symbols $\triangle_{\lambda_{1,2}}$ denote the central 
second-order finite difference
operators of step $\sigma$, 

\begin{equation}
\triangle_{\lambda_{1}} k(\lambda_1,\lambda_2;t)=
k(\lambda_1+\sigma,\lambda_2;t)+
k(\lambda_1-\sigma,\lambda_2;t)-2k(\lambda_1,\lambda_2;t)
\label{second}
\end{equation} 
On the other hand, for  the sites adjacent to the tracer position,
i.e. $(\sigma,0)$, $(-\sigma,0)$ and $(0,\pm\sigma)$, 
respectively,
 we get

\begin{eqnarray} 
\label{(sigma,0)}
\dot{k}(\sigma,0;t)&=&
\frac{l}{4\tau^*}\{\nabla_{\lambda_1}+\triangle_{\lambda_2}\}
k(\sigma,0;t) -\frac{f+g}{\tau^*}k(\sigma,0;t)
+\frac{f}{\tau^*}+\nonumber\\
&+&\frac{1}{\tau}\sum_{(X_{tr},\eta)}(1-\eta(X_{tr}+\sigma,0))
P(X_{tr},\eta;t)\eta(X_{tr}+2\sigma,0),
\end{eqnarray}

\begin{eqnarray} 
\dot{k}(-\sigma,0;t)&=&\frac{l}{4\tau^*}\{\nabla_{\lambda_1}+\triangle_{\lambda_2}\}
k(-\sigma,0;t) -\frac{f+g}{\tau^*}k(-\sigma,0;t)
+\frac{f}{\tau^*}+\nonumber\\
&-&\frac{1}{\tau}\sum_{(X_{tr},\eta)}(1-\eta(X_{tr}+\sigma,0))
P(X_{tr},\eta;t)\eta(X_{tr}-\sigma,0),
\end{eqnarray}
and

\begin{eqnarray} 
\label{(0,+-sigma)}
&&\dot{k}(0,\pm\sigma;t)=\frac{l}{4\tau^*}\{\nabla_{\lambda_2}+\triangle_{\lambda_1}\}
k(0,\pm\sigma;t) -\frac{f+g}{\tau^*}k(0,\pm\sigma;t)
+\frac{f}{\tau^*}+\nonumber\\
&+&\frac{1}{\tau}\sum_{(X_{tr},\eta)}(1-\eta(X_{tr}+\sigma,0))
P(X_{tr},\eta;t)\{\eta(X_{tr}+\sigma,\pm\sigma)-\eta(X_{tr},\pm\sigma)\},
\end{eqnarray}
where $\nabla_{\lambda_{1,2}}$ stand for the forward difference operators, of the form
 
\begin{equation}
\label{first}
\nabla_{\lambda_{1}} k(\lambda_1,\lambda_2;t) = k(\lambda_1+\sigma,\lambda_2;t)- 
k(\lambda_1,\lambda_2;t),
\end{equation}
and
\begin{equation}
\nabla_{\lambda_{2}} k(\lambda_1,\lambda_2;t) = k(\lambda_1,\lambda_2+\sigma;t)- 
k(\lambda_1,\lambda_2;t),
\end{equation} 
Note, that Eqs.(\ref{(sigma,0)}) to (\ref{(0,+-sigma)})
are different from Eq.(\ref{lambda>sigma}), 
since  the 
density profile at the neighboring
 to the tracer sites is strongly perturbed by its
asymmetric hopping rules. As a matter of fact,
 Eqs.(\ref{(sigma,0)}) to (\ref{(0,+-sigma)})
can be thought off as the boundary conditions 
for Eq.(\ref{lambda>sigma}).

Now, several remarks on Eqs.(\ref{lambda>sigma}) to 
(\ref{(0,+-sigma)}) are in order. 
First of all, one notices that these equations are not closed with respect to $k(\lambda_1,\lambda_2)$ but rather 
coupled
to the third-order, tracer-particle-particle
 correlation functions.  In turn, if we attempt to derive the evolution equations for the third-order correlation functions, we find that the latter appear to be coupled to the fourth-order correlations. 
Consequently, in order
to determine the tracer velocity, one faces
the problem of solving an infinite hierarchy of coupled equations
for the correlation functions.  
Here we resort to the simplest non-trivial 
closure of Eqs.(\ref{lambda>sigma}) to 
(\ref{(0,+-sigma)})   
in terms of the pairwise correlation function $k(\lambda_1,\lambda_2)$,
which is based on the following decoupling approximation of the
third-order correlation functions, 

\begin{eqnarray}
&&\sum_{(X_{tr},\eta)} 
\eta(X_{tr}+\lambda_1,\lambda_2)(1-\eta(X_{tr}+\sigma,0))P(X_{tr},\eta;t)\approx\nonumber\\
&\approx&\Big\{\sum_{(X_{tr},\eta)}\eta(X_{tr}+\lambda_1,\lambda_2)P(X_{tr},\eta;t)\Big\}
\times\Big\{\sum_{(X_{tr},\eta)}(1-\eta(X_{tr}+\sigma,0))P(X_{tr},\eta;t)\Big\}=\nonumber\\
&=&k(\lambda_1,\lambda_2;t)(1-k(\sigma,0;t)),
\label{decouplage}
\end{eqnarray}
i.e. the average  with the weight
$P(X_{tr},\eta;t)$ of the product of several occupation numbers of
different sites is set equal to the product of their average values with the weight
$P(X_{tr},\eta;t)$.

We hasten to remark that the approximate closure of the evolution equations 
in Eq.(\ref{decouplage}), as well as some similar but not obviously equivalent
approximations \cite{elliott,tahir}, 
 have been already employed for studying related 
models of tracer diffusion in hard-core lattices gases and 
shown to provide quite 
accurate description of the dynamical and stationary behavior.
The decoupling in Eq.(\ref{decouplage})
 has been first introduced in Ref.\cite{burlatsky}
 to determine the properties of the driven 
tracer diffusion in a
one-dimensional hard-core lattice gas with the conserved number of
particles, i.e.  without exchanges of particles with the reservoir.
Extensive numerical simulations performed 
in Ref.\cite{burlatsky} have demonstrated
that such a decoupling is quite a plausible
approximation for the model under study. 
Moreover,  rigorous
probabilistic analysis of Ref.\cite{olla} has shown 
that for this model the results
 based on the  decoupling scheme in Eq.(\ref{decouplage})
are essentially exact. 
Furthermore, the same closure procedure
has been applied recently to study spreading 
of a hard-core lattice gas from a
reservoir attached to one of the lattice sites \cite{spreading},  and
to treat 
the biased tracer dynamics in
 a one-dimensional model of adsorbed monolayer in contact
with a vapour phase, i.e. a one-dimensional version of the model
to be studied here. Also in these cases an excellent agreement
has been observed between the analytical 
predictions  and Monte Carlo simulations data. Last but not least, 
as we set out to show
elsewhere \cite{ol}, in case of arbitrary force exerted on the tracer
the decoupling  in Eq.(\ref{decouplage}) reproduces
the results of Refs.\cite{elliott} and \cite{tahir} for the
tracer diffusion coefficient in two-dimensional hard-core lattice gases
with a conserved particles number (the limit $f,g = 0$, $f/g = const$
in our case); these results are known to be exact in the limits of
small and large particle densities and provide a very
accurate approximation for the tracer diffusion coefficient in two-dimensional hard-core
lattice gases with arbitrary particle density \cite{kehr}.
 We thus expect that it will render a plausible
description of the tracer dynamics in the 
two-dimensional model under study 
and base our further analysis on this mean-field-type approximation.

Using the approximation in Eq.(\ref{decouplage}), we can rewrite 
Eq.(\ref{lambda>sigma}) in the following closed form

\begin{equation} \label{eq approx}
\dot{k}(\lambda_1,\lambda_2;t)={\sl \tilde{L}} \; k(\lambda_1,\lambda_2;t)
+\frac{f}{\tau^*},
\end{equation}
where the operator ${\sl \tilde{L}}$ is given by

\begin{equation} \label{operator}
{\sl \tilde{L}}= \frac{l}{4\tau^*}\Big\{\triangle_{\lambda_1} + 
\triangle_{\lambda_2}\Big\}
+\frac{1}{\tau}\{1-k(\sigma,0;t)\} \nabla_{\lambda_1} -\frac{f+g}{\tau^*}
\end{equation}    
Further more,  we
find 

\begin{equation}
\dot{k}(\sigma,0;t)= \left . \Big({\sl \tilde{L}} \; k(\lambda_1,\lambda_2;t)\Big)
\right|_{\lambda_1=\sigma,\lambda_{2}=0} +\frac{f}{\tau^*} +\frac{1}{\tau}\{1-k(\sigma,0;t)\}k(\sigma,0;t)+ 
\frac{l}{4\tau^*} k(\sigma,0;t),
\label{limite0}
\end{equation}

\begin{equation}
\dot{k}(-\sigma,0;t)= \left . \Big({\sl \tilde{L}} \; k(\lambda_1,\lambda_2;t)\Big)
\right|_{\lambda_1=-\sigma,\lambda_{2}=0} +\frac{f}{\tau^*} +
\frac{l}{4\tau^*} k(-\sigma,0;t),
\label{limite1}
\end{equation}
and

\begin{equation}
\dot{k}(0,\pm\sigma,0;t)= \left . \Big({\sl \tilde{L}} \; k(\lambda_1,\lambda_2;t)\Big)
\right|_{\lambda_1=0,\lambda_{2}=\pm \sigma} +\frac{f}{\tau^*} +
\frac{l}{4\tau^*} k(0,\pm\sigma;t) 
\label{limite3}
\end{equation}
Equations (\ref{eq approx}) to
(\ref{limite3}) constitute a closed system of
equations, which suffice the computation of the density profiles
and tracer velocity.  Note, however, that these equation are non-linear, since $k(\lambda_1 = \sigma,0)$ enters the prefactor before the gradient term,
which makes
 such a
computation to be quite a  non-trivial problem.  
Below we consider the solution of this system of equations
in the
limit $t\to\infty$.

\section{Stationary solution of the evolution equations}

We turn  now to the limit $t \to \infty$ and suppose
that both the tracer velocity 
$V_{tr}(t)$ and the density profile around 
the tracer attain 
stationary non-zero values $V_{tr}$ and 
$k(\lambda_1,\lambda_2)$, i.e. $V_{tr} = lim_{t\to\infty} V_{tr}(t)$
and 
$k(\lambda_1,\lambda_2) = 
lim_{t\to\infty} k(\lambda_1,\lambda_2;t)$. 

Next, it is expedient to rewrite Eqs(\ref{eq approx}) to 
(\ref{limite3}) in terms of an auxiliary function
$h_{n,m}$:

\begin{equation}
h_{n,m}=k(n\sigma,m\sigma) - \rho_L,
\label{def h}
\end{equation}
which determines the local deviation of the density from the average value
$\rho_L$.  In terms of this auxiliary function Eqs(\ref{eq approx}) to 
(\ref{limite3}) become for $(n,m)\neq(0,0),(\pm 1,0),(0,\pm 1)$

\begin{eqnarray}
(1+{\cal P})h_{n+1,m}+h_{n-1,m}+h_{n,m+1}+h_{n,m-1}-4
(1+\frac{{\cal P}}{4}+\delta)h_{n,m}=0,
\label{eq peclet}
\end{eqnarray}
while $h_{n,m}$ in the immediate vicinity of the tracer obey

\begin{equation}
(1+{\cal P})h_{2,0}+h_{1,1}+h_{1,-1}-(3+4\delta)h_{1,0}+{\cal P}\rho_L=0,
\label{limite1 peclet}
\end{equation}

\begin{equation}
h_{-2,0}-(1+{\cal P})h_{-1,0}+h_{-1,1}+h_{-1,-1}-(3+4 \delta)h_{-1,0}-{\cal P}\rho_L=0,
\label{limite2 peclet}
\end{equation}

\begin{equation}
(1+{\cal P})h_{1,1}+h_{-1,1}+h_{0,2}-(3+4\delta)h_{0,1}=0,
\label{limite3 peclet}
\end{equation}

\begin{equation}
(1+{\cal P})h_{1,-1}+h_{-1,-1}+h_{0,-2}-(3+4 \delta)h_{0,-1}=0,
\label{limite4 peclet}
\end{equation}
where $D_0$
denotes 
the diffusion coefficient of an isolated adsorbed 
particle: 
$D_0=l\sigma^2/4\tau^*$, 
the characteristic parameter ${\cal P} = V_{tr} \sigma/D_{0}$ 
is akin to the so-called  " Peclet " number in 
the hydrodynamics, and we have  
the parameter $\delta = \sigma^2 (f + g)/4 D_{0} \tau^*$. 
The parameter $\delta$ can be also written as $\delta = 
(f+g)/l =(1- \rho_L)^{-1} (g/l)$, where the factor $(g/l)$ 
is an exponential of the difference of the energy
 barriers against lateral diffusion and desorption. Hence,
  $\delta$ compares the relative weights of the adsorption/desorption 
and the diffusion events. Note, that the factor $(g/l)$ is usually 
small for most of realistic experimental situations, such that  $\delta$ is small 
provided that $\rho_{L} \ll 1$. On the other hand, at high adsorbate densities, $\delta$ can attain relatively large values.
As well, the Peclet-type number ${\cal P}$ can also be small or large, depending on the physical situation.
Below we will consider different possible limits for ${\cal P}$ and $\delta$ in order to
elucidate
the asymptotic behavior of the tracer velocity. 

In a standard approach, Eqs.(\ref{eq peclet}) to (\ref{limite4 peclet}) can be solved by
 introducing 
the generating function of the local deviation
$h_{n,m}$, i.e. 
 
\begin{equation}
H(z,w)=\sum_{n=-\infty}^{\infty}\sum_{m=-\infty}^{\infty}h_{n,m}z^nw^m,
\label{def fonction generatrice}
\end{equation}
Multiplying Eqs.(\ref{eq peclet}) to (\ref{limite4 peclet}) by 
$z^n$ and $w^m$ and summing over all $n$ and $m$, we find then that the generating
function
$H(z,w)$
is given explicitly by

\begin{equation}
H(z,w)=-{\sl K}(z,w) \Big\{z+(u-4(1+\frac{{\cal P}}{4}+\delta))+\frac{(1+{\cal P})}{z}\Big\}^{-1},
\label{expression fonction generatrice}
\end{equation}
where $u=w+1/w$, and 

\begin{equation}
{\sl K}(z,w) = \Big( (1+{\cal P})h_{1,0}+{\cal P}\rho_L\Big)(z -1) + 
h_{0,1} (u - 2) + (h_{-1,0}-{\cal P}\rho_L) \frac{(1 - z)}{z}
\label{ka}
\end{equation}
Note, that $H(z,w)$ depends on $w$ only in the combination $w+1/w$, which signifies
that $h_{n,m}$ is, as it could be expected, an even function of $m$. 

Now, 
$h_{n,m}$ can be found directly from Eqs.(\ref{expression fonction generatrice}) and 
(\ref{ka}) through the standard inversion formulae \cite{darboux2}, which amounts, however,
 to computating  rather
complex integrals. Here we will use instead
 a more straightforward approach, 
identifying $h_{n,m}$ from the expansion of $H(z,w)$ into the series
 in Eq.(\ref{def fonction generatrice}). To do this, we first expand
 $H(z,w)$ into a series
in powers of $(1 + \delta + {\cal P}/4)^{-1}$, 
which gives (if $z+\frac{1+{\cal P}}{z}+u<4(1+\frac{{\cal P}}{4}+\delta)$
which is always possible)

\begin{equation}
H(z,w)=  {\sl K}(z,w)
\sum_{i=0}^{\infty}\left(4 (1 + \frac{{\cal P}}{4} + \delta) \right)^{- 
i - 1}\left(z+\frac{(1+{\cal P})}{z}+u\right)^i
\label{sum}
\end{equation}
Next, expanding $(z+(1+{\cal P})z^{-1}+u)^i$ in series in powers of $w$, we rewrite
Eq.(\ref{sum}) 
as a multiple series of the form

\begin{equation}
H(z,w)=  {\sl K}(z,w)
\sum_{i=0}^{\infty}\sum_{k=0}^{\infty}\left(   4 (1 + \frac{{\cal P}}{4} + \delta) 
   \right)^{- i-k-1}
\Big(_k^{i+k}\Big)\left(z+(1+{\cal P})\frac{1}{z}\right)^i\sum_{j=0}^{k}\Big(_j^k\Big)w^{k-2j},
\label{mul}
\end{equation}
where $(_k^{i})$ denote the binomial coefficients. 
Lastly, gathering the terms with equal powers of $z$ and $w$, we find from
Eq.(\ref{mul})
that $h_{n,m}$ is given explicitly by

\begin{equation}
h_{n,m}=\frac{\Big[
- (1 +  {\cal P})  h_{1,0} \nabla_{n} F_{n-1,m}
+  h_{-1,0} \nabla_{n} F_{n,m}  
+  h_{0,1} \triangle_{m} F_{n,m} - 
 {\cal P} \rho_L (F_{n+1,m} - F_{n-1,m})
\Big]}{4 (1 + \delta + {\cal P}/4)},
\label{h en fonction de F}
\end{equation}
where

\begin{equation}
\label{calculdeF}
F_{n,m}= (1 + {\cal P})^{-n/2} \;  P(n,m|\zeta),
\end{equation}
with
\begin{eqnarray}
\label{diffu}
P(n,m|\zeta) &= & 
 \int^{\infty}_0 dt \; e^{ - t} \; I_{n}\Big(\frac{\sqrt{1+{\cal P}} t}{2 (1 + \delta + {\cal P}/4)}\Big) 
\; I_{m}\Big(\frac{t}{2 (1 + \delta + {\cal P}/4)}\Big)= \nonumber\\
&=& \frac{1}{(2 \pi)^2} 
\int^{\pi}_{-\pi} \int^{\pi}_{-\pi} d^2{\bf k} \frac{e^{\Big( - i (k_{x} n +
k_{y} m)\Big)}}{1 - \zeta \lambda({\bf k})},
\end{eqnarray}
in which equation  $I_{n}(x)$ denotes the modified Bessel function,
while the parameter $\zeta$  (which appears to be an important control
parameter and which will be repeatedly
used in what follows) is defined as

\begin{equation}
\zeta = \frac{1 + \sqrt{1 + {\cal P}}}{2 (1 + \delta + {\cal P}/4)},
\end{equation}
and
\begin{equation}
\label{struc}
\lambda({\bf k}) = \frac{cos(k_y) + \sqrt{1 + {\cal P}} cos(k_x)}{1 + \sqrt{1+{\cal P}}}
\end{equation}
Note, that $P(n,m|\zeta)$ can be thought off, in view of the form of Eq.(\ref{diffu}), as the
generating function for the probability of occupancy of the site $(n,m)$ (the lattice Green function)
for a particular type of
two-dimensional P{\'o}lya random walk, whose structure function is defined by 
Eq.(\ref{struc}) (see, e.g. Refs.\cite{weiss} and \cite{hughes} for more details).

Now, two comments on the just-derived expression for $h_{n,m}$, Eq.(\ref{h en fonction de F}), 
are in order. First of all, we note that Eq.(\ref{h en fonction de F}) is not closed, since 
it contains on the right-hand-side
unknown functions; namely,  $h_{\pm 1,0}$ and $h_{0,1}$. Second, 
 all $h_{n,m}$
 are functions of the tracer terminal velocity, which still remains undefined.
 Consequently, in the remainder we have to find
 $h_{\pm 1,0}(V_{tr})$ and $h_{0,1}(V_{tr})$ as functions of the microscopic parameters, 
(such as $f$, $g$, $\tau$ and etc.), and $V_{tr}$, and then, inserting 
the obtained expression for $h_{1,0}(V_{tr})$ to Eq.(\ref{vitesse}), derive a closed with respect 
to $V_{tr}$ equation.

As a matter of fact, the deviations from $\rho_L$ in the immediate vicinity of the tracer, i.e.  $h_{\pm 1,0}(V_{tr})$ and $h_{0,1}(V_{tr})$,  
can be found directly from Eq.(\ref{h en fonction de F}).
 Setting $n=\pm1$, $m=0$ and $n=0$,$m=1$, we find from
 Eq.(\ref{h en fonction de F}) that $h_{\pm1,0}(V_{tr})$ and $h_{0,1}(V_{tr})$ obey
 the following system of linear algebraic equations

\begin{equation} 
\label{system}
\tilde{A}(V_{tr}) \times \tilde{h}(V_{tr}) = {\cal P}\rho_L\tilde{F}(V_{tr}), 
\end{equation}
where the column-matrices $\tilde{h}(V_{tr})$ and $\tilde{F}(V_{tr})$  are defined as

\begin{equation}
\label{col}
\tilde{h}(V_{tr}) =   
\begin{pmatrix}
h_{1,0}\\
h_{-1,0}\\
h_{0,1}
\end{pmatrix}
 \;\;\;\;
\tilde{F}(V_{tr})
=
\begin{pmatrix} 
F_{2,0} -  F_{0,0}\\
F_{0,0} -  F_{-2,0}\\
F_{1,1} -  F_{-1,1},
\end{pmatrix}
\end{equation}
while the square matrix $\tilde{A}(V_{tr})$ has the form

\begin{equation}
\label{A}
\tilde{A}(V_{tr}) = 
\begin{pmatrix}
-(1+{\cal P})(\alpha_{\zeta}
+ \nabla_{n} F_{0,0})&&\nabla_{n} F_{1,0} && \triangle_{m} F_{1,0}\\
         -(1+{\cal P}) \nabla_{n} F_{-2,0}&&- \alpha_{\zeta}  + \nabla_{n} F_{-1,0}&&\triangle_{m} F_{-1,0}\\
          -(1+{\cal P}) \nabla_{n} F_{-1,1}&&\nabla_{n} F_{0,1}&&- \alpha_{\zeta}  + 
\triangle_{m} F_{0,1},
\end{pmatrix} 
\end{equation}
where $\alpha_{\zeta} = 2(1+\sqrt{1 + {\cal P}})/\zeta$ 
and $\nabla_{n}$ ($\triangle_{m}$) are the finite difference operators of unit step 
in Eq.(\ref{second}) ((\ref{first})), acting on the variable $n$
($m$). To define the tracer terminal velocity, we merely have to know
$h_{1,0}(V_{tr})$. It follows from Eqs.~(\ref{system}) to (\ref{A})
that $h_{1,0}(V_{tr})$ is given by

\begin{equation}
\label{h}
h_{1,0}(V_{tr}) =  {\cal P}  \rho_L \; \frac{\rm{det } \; \tilde{h}_{1,0}(V_{tr})}{\rm{det}  
\; \tilde{A}(V_{tr})},
\end{equation}
where $\tilde{h}_{1,0}(V_{tr})$ is the matrix obtained from (\ref{A})
by replacing the first column by $\tilde{F}(V_{tr})$.

Now, 
we are in position to obtain a closed equation, which determines implicitly the tracer particle 
terminal velocity in terms of the parameters $f$, $g$, $\rho_L$ and $D_0$. Substituting 
Eqs.(\ref{h})
 and (\ref{def h})
into Eq.(\ref{vitesse}), we find that $V_{tr}$ obeys

\begin{equation}
\label{speed}
V_{tr} = \frac{\sigma (1 - \rho_L)}{\tau} \; \Big[ 1 + \frac{\sigma^2 \rho_L}{D_{0} \tau}
\; \frac{\rm{det } \; \tilde{h}_{1,0}(V_{tr})}{\rm{det}  \; \tilde{A}(V_{tr})}\Big]^{-1},
\end{equation} 
where the first term is the mean field result for the tracer velocity (see
Eq.~(\ref{mean-field})), obtained under assumption that the monolayer
is perfectly stirred; terms in brackets stem from non-linear
cooperative effects associated with formation of inhomogeneous density 
profiles, whose characteristic properties depend parametrically on the 
tracer velocity. 
Equation (\ref{speed}), which is the main general analytical result of our paper,
is certainly too complicated  to be 
solvable in the general case. Below we will study some 
of its asymptotical  solutions.

\section{Asymptotic behavior of the tracer 
stationary velocity and  density profiles at a fixed distance from the tracer}

In this section we consider the asymptotical behavior of the
tracer particle velocity  and  density profiles
at a fixed distance from the tracer  in the limit of
slow and fast particles diffusion ($n$ and $m$ are taken arbitrary but fixed, while $D_{0}$ is allowed to vary).  
As well, we will also specify different regimes corresponding
to the cases with ${\cal P} \gg 1$ and ${\cal P} \ll 1$, which may be realized at different monolayer densities. 
We note that such an analysis turns out to be rather non-trivial, which happens
not only because of the non-linearity of the
equations involved, but chiefly due to the non-uniformity of the asymptotical expansions with respect to the parameter
$\delta$. As a matter of fact, such a non-uniformity is just the consequence of the qualitatively
 different behavior
occurring in the system with explicitly 
conserved ($\delta = 0$) and non-conserved  ($\delta > 0$)  particles number.
 Consequently, the results for the system without adsorption and desorption can not  be obtained as a limit $\delta
\to 0$ of the corresponding results with arbitrary (even vanishingly
small) values of $\delta$.  Lastly, we note that the limits $D_{0} \to \infty$ (or $D_{0} \to 0$) and $n,m \to \infty$
do not commute; hence, we consider the asymptotical behavior of the density profiles at large distances from the tracer separately in Section
6.

\subsection{Limit of a small particles diffusivity.\label{sec:limit-small-part}}

Consider first the asymptotical behavior of  $V_{tr}$
in the limit of a vanishingly small diffusion coefficient $D_{0}$
of the monolayer particles,
$D_{0} \to 0$, which situation can be thought off as a weakly perturbed original
Langmuir model ($D_{0} = 0$). It appears, as we have already remarked, 
 that this asymptotical behavior can
 be different in the case when the particles number
 in the monolayer is not  explicitly conserved, i.e. 
when both $f,g > 0$, and in the case with a conserved particles
 number, when both $f,g = 0$ but their ratio is fixed,  $f/g = \rho_L/(1 - \rho_L)$. 
Consequently, we have to study these two cases separately.  

{\it A. Non-conserved particles number}.  We assume first that both $f$ and  
$g$ are non-zero, such that $\delta \neq 0$, 
which means that adsorption and
 desorption do change
 the occupation of each site. In this case, it seems natural  to
 expect that the tracer will never be blocked and will always
 continue moving at a constant velocity.
In other words, we anticipate that in this case
the
 velocity attains a constant non-zero value 
$V^{(0)} > 0$ as $D_{0} \to 0$ and hence, that 
${\cal P} \gg 1$, $\delta \gg 1$, which means that  
$\zeta \ll 1$. We will refer to this limit of $D_0 \to 0$, $\delta \gg 1$ and ${\cal P} \gg 1$
as the {\it limit (a)}. 
We seek then
$V_{tr}$ in the form

\begin{equation}
\label{exp}
V_{tr} = V^{(0)} + V^{(1)} \Big(\frac{D_{0} \tau^*}{\sigma^2}\Big) + 
{\mathcal O} \Big((\frac{D_{0} \tau^*}{\sigma^2})^2\Big).
\end{equation}

Further on, expanding the denominator 
in Eq.(\ref{diffu}) in powers of the parameter $\zeta$, 
 we calculate to the first 
order in powers of $D_{0}\tau^*/\sigma^2$ the functions  $ F_{n,m}$ involved in
Eq.(\ref{speed}). These are listed in Appendix 1. Then,
substituting the obtained expressions for $F_{n,m}$
into Eq.(\ref{speed}), we have then

\begin{equation}
\label{V0}
V^{(0)} = \frac{(1-\rho_L) \sigma }{\tau} \left\{1 + \frac{\rho_L \tau^* }{(f+g)\tau} \right\}^{-1}
\end{equation}
and

\begin{equation}
\label{V1}
V^{(1)} = \rho_L(1-\rho_L)\sigma\frac{\rho_L+2+3(f+g)\tau/\tau^*}{\tau^*(1+(f+g)\tau/\tau^*)(\rho_L+(f
+g)\tau/\tau^*)^2}
\end{equation}

Further on, using Eqs.(\ref{V0}) and (\ref{V1}), we find the following explicit
results for the deviations from the equilibrium mean density 
$\rho_{L}$ in the immediate vicinity
of the tracer.  These obey, 
in the {\it limit (a)}:

\begin{equation} 
h_{1,0}=\frac{\rho_L (1 - \rho_L)}{\rho_L + (f + g) \tau/\tau^*} 
\left\{1-\frac{\tau}{\tau^*} \frac{(\rho_L+2+3(f+g)\tau/\tau^*)}{\Big(1+(f+g)\tau/\tau^*\Big)\Big(\rho_L+(f+g)\tau/\tau^*\Big)} \Big(\frac{D_{0} \tau^*}{\sigma^2}\Big)\right\} + 
{\mathcal O}\Big((\frac{D_{0} \tau^*}{\sigma^2})^2\Big),
\label{eq:18}
\end{equation}

\begin{eqnarray}
\label{h-10}
h_{-1,0} = - \frac{\rho_L (1 - \rho_L)}{1 + (f + g) \tau/\tau^*} \left\{ 1 -
\frac{\tau}{\tau^*}\frac{(\rho_L+2+3(f+g)\tau/\tau^*)}{\Big(1+(f+g)\tau/\tau^*\Big)^2}
\Big(\frac{D_{0} \tau^*}{\sigma^2}\Big) \right\} +
 {\mathcal O}\Big((\frac{D_{0} \tau^*}{\sigma^2})^2\Big),
\end{eqnarray}
and 

\begin{equation}
\label{h011}
h_{0,1} = \frac{\rho_L (1 - \rho_L)^2}{ (f+g)(1 + (f + g)\tau/\tau^*)^2} 
\Big(\frac{D_{0} \tau^*}{\sigma^2}\Big) + {\mathcal O}\Big((\frac{D_{0} \tau^*}{\sigma^2})^2\Big),
\end{equation}
 As it could be expected intuitively, $h_{1,0} > 0$ and $h_{-1,0} < 0$, which means 
that there is a condensed, "jammed" region in front of the tracer, and a depleted region past the
tracer. On the other hand, the fact that $h_{0,1} > 0$ and that $h_{\pm 1,0}$ under
 certain conditions are non-monotonous functions of $\rho_L$ appear to be non-trivial.
 We will  discuss 
 the characteristic properties of these two regions at the end of this 
section and in the section \ref{sec:Asympt-forms-dens}.

Note, that the result in Eq.(\ref{V0}) can be simply obtained from the initial system
of equations (\ref{eq peclet}) to (\ref{limite4 peclet}) by 
 assuming that the deviation
$h_{n,0}$ 
from the mean density  is  zero for $n\geq 2$, i.e. the density profile
in front of the density profile is an abrupt
step function.
Then, it follows from Eqs.(\ref{limite1 peclet}) and (\ref{limite2 peclet}) that
$h_{\pm 1,0} \approx  \pm  {\cal P} \rho_L/(3+4 \delta)$ 
Plugging this approximate result into Eq.(\ref{vitesse}) and turning to the limit $D_{0} \to 0$
we actually recover
Eq.(\ref{V0}).  
Correction term in Eq.(\ref{V1}) is thus associated with
the appearance of a smooth, long-range density variation with the distance from the
tracer due to diffusion, which couples effectively 
the evolution of $h_{n,m}$ at different
lattice sites.  Note finally, that Eq.(\ref{V0}) reduces to the essentially mean-field in Eq.(\ref{mean-field}), i.e. 
$V_{tr} = \sigma (1 - \rho_L)/\tau$, in the limit $\tau^* \to 0$, i.e. in the limit
when adsorption/desorption processes proceed at  infinitely fast rate resulting in a 
complete homogenization of the monolayer.

Next, using Eq.~(\ref{h en fonction de F}), and taking advantage of the fact that
to the lowest non-trivial
order in powers of $D_{0}$ the function $F_{n,m}$ in 
Eq.(\ref{calculdeF}) is given by

\begin{equation}
\label{F}
F_{n,m} 
\sim \Big(_{|n|} ^{|n|+|m|}\Big) \Big(\frac{D_{0}}{V^{(0)} \sigma + \sigma^2 (f +
g)/\tau^*}\Big)^{|n| + |m|} J_{n},
\end{equation}
where

\begin{equation}
J_{n} = \left\{\begin{array}{ll}
1,     \mbox{ if $n > 0$,} \nonumber\\
(V^{(0)} \sigma/D_{0})^{|n|},     \mbox{ if $n < 0$,}
\label{eq:14}
\end{array}
\right.
\end{equation}
we find that, at a finite distance from the tracer, the local deviation from the equilibrium mean density obey:
\begin{equation}
h_{n,m}\sim\frac{\rho(1-\rho)}{\rho+(f+g)\tau/\tau^*}\Big(_{n-1}^{n-1+|m|}\Big)
\left[\frac{D_0\tau^*}{\sigma^2(f+g)}\frac{\rho+(f+g)\tau/ \tau^*}{1+(f+g)\tau/
\tau^*}\right]^{n-1+|m|}\mbox{  
for $n>0$,}
\label{eq:15}
\end{equation}
\begin{equation}
h_{n,m}\sim-\frac{\rho(1-\rho)}{1+(f+g)\tau/\tau^*}\Big(_{|n|}^{|n|+|m|}\Big)
\left[\frac{1-\rho}{1+(f+g)\tau/ \tau^*}\right]^{|n|-1}
\left[\frac{D_0\tau^*}{\sigma^2(f+g)}\frac{\rho+(f+g)\tau/ \tau^* }{1+(f+g)\tau/
\tau^*}\right]^{|m|}\mbox{  
for $n<0$,}
\label{eq:16}
\end{equation}
and
\begin{equation}
h_{0,m}\sim\frac{|m|\rho(1-\rho)^2}{(1+(f+g)\tau/ \tau^*)( \rho +(f+g)\tau/
\tau^*)}\left[\frac{D_0\tau^*}{\sigma^2(f+g)}\frac{ \rho +(f+g)\tau/
\tau^*}{1+(f+g)\tau/ \tau^*}\right]^{|m|},
\label{eq:17}
\end{equation}
Note, that these expressions generalize the leading
order contributions of Eqs.(\ref{h10})-(\ref{h011}). It may be also worthy to remark that
$h_{n,m}$ turns out to be  a non-separable
function with respect to the variables $n$ and $m$. The density
profiles defined by Eqs. (\ref{eq:15}) to (\ref{eq:17}) are depicted
in Fig.2a.

Lastly, we comment on the range of applicability of all the results of 
this subsection,  which 
are based on the assumption that ${\cal P} \gg 1$ and $\zeta \ll 1$.
This implies, in turn, that these results are only valid
when
$D_{0}$ obeys the inequality

\begin{equation}
\label{ineq}
D_{0}  \ll 
\frac{\sigma^2 (1 - \rho_L)}{2 \tau} 
\left\{1 + \frac{\rho_L \tau^* }{(f+g)\tau} \right\}^{-1},
\end{equation}
which sets up an upper bound on $D_0$ showing how 
small it should be to insure the utility the previous expressions.

{\em B. Conserved particles number.} Consider next the special case when both $f,g \to 0$,
 but their ratio 
$f/g$ is kept fixed, $f/g = \rho_L/(1-\rho_L)$, which  corresponds
to the situation with a conserved number of particles (no adsorption/desorption),
i.e. a standard model of a two-dimensional hard-core lattice gas.
In this case the parameter $\delta$ is exactly equal to zero,  $\delta = 0$, 
 $V^{(0)}$ should vanish according to Eq.(\ref{V0}), 
and hence, $V_{tr}$ is expected to
scale linearly with $D_{0}$
in the
lowest order with respect to the particle diffusion coefficient. Note, however, that in the case $\delta =
0$ the inequality in Eq.(\ref{ineq}) is not satisfied for any
 $D_{0}$ except $D_0 = 0$, which means that, in principle, Eqs.(\ref{V0}) and (\ref{V1}) do not describe
 the behavior of the tracer velocity properly.

Let us now examine separately the case $\delta = 0$ for which the results
(\ref{V0}) to (\ref{h011}) do not apply. 
Assume first that the Peclet number is large, i.e. ${\cal P} \gg 1$, which will be
checked for consistency afterwards. Then, in this limit $D_0 \to 0$, $\delta = 0$ and ${\cal P} \gg 1$, which
will be referred to in the remainder as the {\it limit (b)}, we have that $\zeta \ll 1$ and we can use the
expressions for $F_{n,m}$ (see Appendix 1) with $\delta=0$,
provided that
 $V^{(0)}$ is  replaced by $V_{tr}$. This yields for the terminal
velocity the following result
\begin{equation}
\label{VI}
V_{tr} \sim \frac{2 (1 - \rho_L)}{\sigma \rho_{L}} D_{0}
\end{equation}
It follows then from Eq.(\ref{VI}) that here the Peclet number is given approximately by ${\cal P} \approx 2 (1 -
\rho_L)/\rho_L$;  hence, Eq.(\ref{VI}) is consistent with the underlying assumption only
when $\rho_L \ll 1$. In this limit, we find that
$h_{1,0} \sim 1, \;\;\;\;\;\;\; h_{-1,0} \sim - \rho_L, \;\;\;\;\mbox
 {     and }\;\;\;\; h_{0,1} \sim \rho_L/2.$
Note that in this case, $h_{0,1}$ is independent of $D_0$ to leading
order, contrarily to the prediction of Eq.~(\ref{h011}).

More generally, using Eq.~(\ref{h en fonction de F}) and
Eqs.(\ref{eq:1})-~(\ref{eq:6}), we find that at a finite distance from 
the tracer
\begin{equation}
h_{n,m}\sim\Big(_{n-1}^{n-1+|m|}\Big)\Big(\frac{\rho}{2}\Big)^{n-1+|m|}\mbox{ for 
$n>0$,}
\label{eq:19}
\end{equation}
\begin{equation}
h_{n,m}\sim-\Big(_{|n|}^{|n|+|m|}\Big)\frac{|m|}{|m|-n+1}\Big(\frac{\rho}{2}\Big)^{|m|}\mbox{ 
for 
$n<0$ and $m\neq0$}
\label{eq:20}
\end{equation}
and
\begin{equation}
h_{0,m}\sim|m|\left(\frac{\rho}{2}\right)^{|m|}.
\label{eq:21}
\end{equation}
Furthermore, using the expansion 
\begin{equation}
F_{n,0}=1+\frac{2(n+1)}{{\cal P}}+{\mathcal O}\Big(\frac{1}{{\cal P}^2}\Big) \mbox{
for $\delta=0$ and $n<0$,} 
\label{eq:22}
\end{equation}
we have
\begin{equation}
h_{n,0}\sim-\rho\mbox{ for $n<0$}
\label{eq:23}
\end{equation}
Again, these expressions are not separable in $n$ and $m$. The density 
profiles defined by Eqs.~(\ref{eq:19}) to (\ref{eq:23}) are depicted
in Fig.2b.

Let us consider now the {\it limit (c)} when $D_{0} \to 0$, $\delta = 0$, but ${\cal P} \ll 1$. In this limit
we evidently have  $\zeta \sim  1$. 
Notice now that for $\zeta = 1$,
 the integral in Eq.(\ref{diffu}) diverges at small values of the
wave-vector ${\bf k}$, so that when $1 - \zeta$ is small 
but non-zero, the integral over ${\bf k}$ is dominated by the small-${\bf
k}$ behavior (strictly speaking, this is {\it a priori} true only if $n$ and $m$ are
large enough to insure the validity
 of the stationary phase method) . Following Montroll and Weiss \cite{weiss}, who have studied
 asymptotic behavior
of $P(n,m|\zeta)$ in Eq.(\ref{diffu}) 
within the context of P{\'o}lya random walk on 2D lattices,
(see also Ref.\cite{hughes}),
we first represent the structure function in Eq.(\ref{struc}) as
$1 - \lambda({\bf k}) \sim \frac{1}{2} \Big(\sigma_{x}^2 k_{x}^2 + \sigma_{y}^2
k_{y}^2 \Big)$,
where 

\begin{equation}
\label{scales}
\sigma_{x} = \frac{(1 + {\cal P})^{1/4}}{(1 + \sqrt{1+{\cal P}})^{1/2}}; \; \sigma_{y} = \frac{1}{
(1 + \sqrt{1+{\cal P}})^{1/2}}
\end{equation}
Then, using the stationary phase method, one finds \cite{hughes}
 the following representation
of $P(n,m|\zeta)$ in Eq.(\ref{diffu}), which holds 
in the limit $\zeta \to 1$: 

\begin{equation}
\label{m}
P(n,m|\zeta) \sim \frac{1}{\pi \sigma_{x} \sigma_{y}} K_{0}\Big(\Big[2 (1 - \zeta) (\frac{n^2}{\sigma_{x}^2} + 
\frac{m^2}{\sigma_{y}^2})\Big]^{1/2}\Big),
\end{equation}
where $K_{0}(z)$ is the modified Bessel function of the third kind. In
the limit $\zeta \to  1$, Eq.~(\ref{m}) yields:

\begin{equation}
\label{as}
P(n,m|\zeta) =- \frac{2}{\pi} ln(\frac{{\cal P}}{4}) 
- \frac{ln(n^2 + m^2)}{\pi} - \frac{2 \gamma}{\pi} 
 + {\mathcal O}({\cal P} ln{\cal P})
\end{equation}

Note that Eq.~(\ref{as}) only holds at intermediate distances from the 
origine, since, on the one hand, $n$ and $m$ have to be large enough
in order to apply th stationary phase method, but, on the other hand,
the argument of  $K_0$ in Eq.~(\ref{m}) should be small.  
However, 
as shown in Ref.\cite{mccrea}, the asymptotical result in Eq.(\ref{as}) 
  serves   as a very
good estimate when $n$ and $m$ are  sufficiently large
(in practical, as soon as $(n^2+m^2)^{1/2}\geq3$). 
Since we are also interested
to know  $h_{\pm 1,0}$ and  $h_{0,\pm 1}$ precisely,  we list below 
some particular values of $P(n,m|\zeta)$ in the immediate vicinity of the
tracer. Exact values of $P(n,m|\zeta\to1)$ are presented in
Ref.\cite{mccrea}. From these, we find  

\begin{equation}
\nabla_{n} F_{0,0}=-1+ {\mathcal O}({\cal P} ln({\cal P})), \;\;\;\nabla_{n} F_{1,0}=-3+8/\pi+
{\mathcal O}({\cal P} ln({\cal P})),\;\;\; 
\triangle_mF_{1,0}=2-8/\pi+ {\mathcal O}({\cal P} ln({\cal P}))  
\label{eq:33}
\end{equation}
\begin{equation}
\nabla_{n} F_{-2,0}=3-8/\pi+{\mathcal O}({\cal P} ln({\cal P})),\;\;\;\nabla_{n} F_{-1,0}=1+
{\mathcal O}({\cal P} ln({\cal P})),\;\;\;
\triangle_mF_{-1,0}=2-8/\pi+ {\mathcal O}({\cal P} ln({\cal P}))
\label{eq:34} 
\end{equation}
\begin{equation}
\nabla_{n} F_{-1,1}=-1+4/\pi+{\mathcal O}({\cal P} ln({\cal P})),\;\;\;\nabla_{n}
F_{0,1}=1-4/\pi+{\mathcal O}({\cal P} ln({\cal P})),\;\;\;
\triangle_mF_{0,1}=-2+8/\pi+{\mathcal O}({\cal P} ln({\cal P})),  
\end{equation}
which allow us to determine the local deviations in the immediate vicinity of the tracer. 
We find that in the {\it limit (c)} 
 the local deviations are to the leading   order in ${\cal P}$:

\begin{equation}
\label{d}
h_{1,0} = \alpha \rho_L {\cal P}  + {\mathcal O}({\cal P}^2)\;\;\;\; 
h_{-1,0} = - \alpha \rho_L {\cal P}  +{\mathcal O}({\cal P}^2)\;\;\;\; 
h_{0,1} =  {\mathcal O}({\cal P}^2),
\end{equation}
where $\alpha$ is a numerical constant, 

\begin{equation}
\alpha = - \frac{ P(2,0|1)-P(0,0|1)}{4 + P(2,0|1)-P(0,0|1)} = \frac{\pi-2}{2} \approx 0.571
\label{eq:32}
\end{equation}

Finally, inserting Eqs.(\ref{d}) to Eq.(\ref{vitesse}), we find the following
 expression for the tracer terminal velocity in the limit (c):

\begin{equation}
\label{vi}
V_{tr} 
= \frac{(1 - \rho_L) D_0}{\alpha \sigma \rho_L} \Big[ 1 - 
\frac{D_0 \tau}{\alpha \sigma^2 \rho_L} + {\mathcal O}\Big((\frac{D_0
\tau^*}{\sigma^2})^2\Big)\Big]
\end{equation}

Note, that similarly to Eq.(\ref{VI}), in the limit (c) the tracer velocity
 $V_{tr}$ vanishes in proportion to $D_0$;
the prefactor is, however, different from that in Eqs.(\ref{V1}) and Eq.(\ref{VI}).  
The Peclet number in this case is again ${\cal P} \sim (1 - \rho_L )/\rho_L$, which means, in view of the
consistency with the initial assumption, that the limit (c) is only realized and 
Eq.(\ref{vi}) is only valid when $\rho_{L} \sim 1$.
 We also remark,  
that  
a similar to Eq.(\ref{vi}) expression for the tracer velocity with, however, 
 a bit different
value of $\alpha$, has been obtained prior in \cite{deconinck}, 
which studied the stationary velocity of a driven tracer particle 
in a two-dimensional lattice gas with a conserved particles number.

Lastly, using Eq.~(\ref{h en fonction de F}) and
Eq.~(\ref{as}), we find that at a finite, but large enough distance from the tracer, 
\begin{eqnarray}
h_{n,m}
&\sim&\frac{(1-\rho)(1+\alpha)}{4\pi\alpha}\ln\left(\frac{(n+1)^2+m^2}{(n-1)^2+m^2}\right),
\label{eq:26}
\end{eqnarray}
which is again a non-separable function of variables $n$ and
$m$. Profiles defined by Eq.~(\ref{eq:26}) are depicted in Fig.2c.

\subsection{Limit of a large particles diffusivity.}

We
turn next to the analysis of the asymptotical behavior of $V_{tr}$ in the limit of very fast
monolayer particles diffusion supposing that both $f$ and $g$ are
non-zero, which will be called  the {\it limit (d)}.
 Clearly, when $D_0 = \infty$, one should recover for $V_{tr}$
the trivial mean-field result in Eq.(\ref{mean-field}). We search, hence, for the
first correction to this result, representing the tracer velocity as

\begin{equation}
\label{Vfast}
V_{tr} = \frac{\sigma (1 - \rho_L)}{\tau} + V_{(1)} \Big(\frac{\sigma^2}{D_0 \tau}\Big) + 
{\mathcal O}\Big((\frac{\sigma^2}{D_0 \tau})^2\Big)
\end{equation}

Now, one can readily notice that in this case we deal, in essence, with a situation which is
quite similar to the {\it limit (c)} studied in the previous subsection. Namely, 
both $\delta$ and ${\cal P}$ are small and hence, 
$\zeta$ should be of order of unity, $\zeta \sim
1$. Consequently, all the analysis of the last subsection applies 
here, except that $\zeta$ takes a bit different form.  

Noticing next that here $\zeta \approx 1 - \delta$ and
 calculating the integral in Eq.(\ref{diffu}) along essentially the same lines
as it was done in the previous subsection, we find that in the limit of fast particles diffusion
$P(n,m|\zeta \to 1)$ is given by :

\begin{equation}
P(n,m|\zeta \to 1) \sim - \frac{2}{\pi} ln(\delta)
- \frac{ln(n^2 + m^2)}{\pi} - \frac{2 \gamma}{\pi} 
 +{\mathcal O}(\frac{ln(D_0)}{D_0^2}),
\end{equation}
which expansion  holds for large enough values of the variables
$n$ and $m$.  Consequently, we find that in the limit (d) the local deviations from the equilibrium
 mean density in the immediate
vicinity of the tracer obey Eqs.(\ref{d})  and hence,
the first correction to Eq.(\ref{mean-field}) reads 

\begin{equation}
\label{vel}
V_{(1)} = - \rho_{L} (1 - \rho_{L}) \frac{\alpha \sigma}{\tau}
\end{equation}

As it could be expected intuitively, 
$V_{(1)}$ appears to be
negative, i.e. the actual tracer terminal velocity is
 lower than that given by Eq.(\ref{mean-field}). Note also that
$V_{(1)}$ is a non-monotonous function of $\rho_L$  and is maximal when $\rho_L = 1/2$. 
We finally remind that 
the expansions in Eq.(\ref{Vfast}) and Eq.(\ref{vel}) hold, 
by definition, in the limit $D_{0} \to \infty$, which
means they only apply when
both ${\cal P}$ and $\delta$ are much less than unity.

 We remark  that Eq.(\ref{vel}) and Eqs.(\ref{d}) still hold
when $f,g = 0$. On the other hand, it will be  shown in the Appendix that in the
{\it limit (d)} 
the amplitudes and the characteristic lengths
of the density profiles attain different forms depending whether $\delta$
is equal to zero or not. 

Lastly, the form of the density profiles at a finite distance from the tracer is very 
similar to the one found in the limit (c). We have that, at finite but
large enough distance, $h_{n,m}$ is given by: 
\begin{eqnarray}
h_{n,m}
&\sim&\frac{\rho(1-\rho)(1+\alpha)}{4\pi}\frac{\sigma^2}{\tau D_0}\ln\left(\frac{(n+1)^2+m^2}{(n-1)^2+m^2}\right),
\label{h10}
\end{eqnarray}

The remarks following Eq.(\ref{eq:26}) still hold.

\section{Asymptotic forms  of the density profiles at large distances from the tracer.}\label{sec:Asympt-forms-dens}

In this section we consider
the asymptotical behavior of the
  density profiles at large distances  from
 the stationary moving tracer particle. The first subsection deals
with the simple special case $D_0=0$.
The second subsection
will be devoted to the analysis of the 
asymptotic forms of $h_{n,0}$ in the limit 
$n
\to  \infty$, which will be based on our Eqs.(\ref{h en fonction de F}) to (\ref{diffu}).
 Further on, in the next
subsection we will examine a more 
complicated question of the decay of $h_{n,0}$ at large separations past the 
tracer, where we again 
 discuss separately the behavior in the cases
 with non-conserved and conserved particles numbers. Actually, we observe here a very
spectacular effect of a qualitative change of the asymptotical behavior;
the density relaxation to the average value $\rho_L$ proceeds in the former case
exponentially, while in the latter case it is described by a slow algebraic function of
the distance.
Lastly,
 we will study the forms of some integral characteristic
of the density profile, such as, for instance,
 the global deviation of the density from the equilibrium value in the domains $n >
1$ and $n < -1$.

Asymptotic behavior
 of $h_{n,0}$ as $n \to \pm \infty$ can be most conveniently studied
if we introduce the generating function of $h_{n,0}$ of the form

\begin{equation}
\label{h(z)}
h(z) = \sum_{-\infty}^{+\infty}h_{n,0}z^n
\end{equation}

Setting then $m=0$ in Eq.(\ref{h en fonction de F}) and summing over all $n$, we 
find that $h(z)$ is given explicitly by

\begin{equation}
h(z)
=\frac{z(z-1)\Big((1
+ {\cal P}) h_{1,0} + {\cal P}
\rho_L\Big)+(1-z)\Big(h_{-1,0}
- {\cal P} \rho_L\Big)}{\sqrt{(z-z_1)(z-z_2)(z-z_3)(z-z_4)}}- h_{0,1} \sqrt{\frac{(z-z_2)(z-z_3)}{(z-z_1)(z-z_4)}}
\label{fg resommee}
\end{equation}
where the roots $z_i$ are as follows:

\begin{eqnarray}
z_1 = \frac{1 + \sqrt{1 + {\cal P}} + \zeta}{\zeta} - \sqrt{\Big(\frac{1 + \sqrt{1 + {\cal P}}
 + \zeta}{\zeta}\Big)^2 - (1 + {\cal P})}; \nonumber\\
z_2 = \frac{1 + \sqrt{1 + {\cal P}} - \zeta}{\zeta} - 
\sqrt{\Big(\frac{1 + \sqrt{1 + {\cal P}} - \zeta}{\zeta}\Big)^2 - (1 + {\cal P})}; \nonumber\\
z_3 = \frac{1 + \sqrt{1 + {\cal P}} - \zeta}{\zeta} + \sqrt{\Big(\frac{1 + \sqrt{1 + {\cal P}}
 - \zeta}{\zeta}\Big)^2 - (1 + {\cal P})}; \nonumber\\
z_4 = \frac{1 + \sqrt{1 + {\cal P}} + \zeta}{\zeta} + \sqrt{\Big(\frac{1 +
 \sqrt{1 + {\cal P}} + \zeta}{\zeta}\Big)^2 - (1 + {\cal P})}
\label{eq:8}
\end{eqnarray}
and obey
\begin{equation}
0<z_1\leq z_2\leq1<z_3\leq z_4,
\label{ordre des racines}
\end{equation}
such that $h(z)$ is analytic in 
the annular region of inner radius $z_2$
and outer radius $z_3$. 
 This implies
  that both $z_1$ and $z_4$ are irrelevant 
to the large scale behavior and 
the dominant contribution to the asymptotic 
form of $h_{n,0}$ as $n \to \pm \infty$ comes from the 
the behavior of $h(z)$ as $z \to z_2$ or $z \to z_3$. More precisely, the asymptotic form 
$h_{n,0}$ as $n \to  +\infty$ stems from the behavior of $h(z)$ in the vicinity of $z = z_3$, while 
$h_{n,0}$ as $n \to  - \infty$   is associated with the corresponding
behavior near $z = z_2$.

\subsection{Density profiles in the special case $D_0=0$.}

First of all, we  study the special case $D_0=0$, for which the
roots $z_1$ and $z_2$ collapse to the same value
$(1-\rho_L)/(1+(f+g)\tau/\tau^*)$. 
In this special case, which 
corresponds to the original Langmuir model, it is possible to
determine  the density profile exactly. Indeed, here 
Eq.~(\ref{fg resommee}) reduces to
\begin{eqnarray}
h(z)&=&h_{1,0}\frac{(z-1)(z+h_{-1,0}/h_{1,0})}{z(1-z_2/z)}\nonumber\\
&=&h_{1,0}\Big(z-(1-h_{-1,0}/h_{1,0})+h_{-1,0}/h_{1,0}\frac{1}{z}\Big)\sum_{n=0}^{\infty}\left(\frac{z_2}{z}\right)^n\nonumber\\
&=&h_{1,0}\Big(z+\sum_{n=1}^\infty\frac{(1-z_2)z_2^{n-1}}{z^n}\Big),
\label{eq:9}
\end{eqnarray}
which implies
\begin{equation}
h_{n,0}=0,\mbox{ for }n>1\;\;
,h_{n,0}=h_{-1,0}\left(\frac{1+(f+g)\tau/\tau^*}{1-\rho_L}\right)^n,\mbox{
for } n<0, 
\end{equation}
where $h_{1,0}$ and $h_{-1,0}$ are given to the leading order by
 Eqs.(\ref{h10}) and (\ref{h-10}) respectively. These expressions confirm the step
 shaped profile in front of the tracer particle (as anticipated in
 section \ref{sec:limit-small-part}), such that the only perturbed sites are the sites
visited by  the tracer particle, and the  site just in front of
it. These expressions can be also  readily recovered by considering
the limiting form of the equations (\ref{eq peclet})-~(\ref{limite4 peclet}) when $D_0=0$.

\subsection{Asymptotic forms of the density profiles at large distances 
in front of the stationary moving tracer.}

From now on, we assume that the diffusion coefficient $D_0$ is
not equal to zero, and
   we deduce the asymptotical behavior 
of $h_{n,0}$ in the limit $n \to \pm \infty$
from the analysis of the singularities 
of the generating function $h(z)$. 
One has that in the vicinity of $z = z_3$ the generating function $h(z)$ obeys 

\begin{equation}
h(z)=
\frac{z_3(z_3-1)\Big((1
+ {\cal P}) h_{1,0} + {\cal P}
\rho_L\Big)+
(1-z_3)\Big(h_{-1,0}
- {\cal P} \rho_L\Big)}{\sqrt{\left(z_3-z_1\right)\left(z_3-z_2\right)
\left(z_4-z_3\right)}}\frac{1}{\sqrt{z_3-z}}+{\mathcal O}(\sqrt{z_3-z})
\end{equation}
Next, following Darboux method \cite{darboux2} or singularity
analysis of generating functions \cite{darboux1}, we find that 
$h_{n,0}$ in the limit $n \to \infty$ 
follows 

\begin{equation}
h_{n,0} \sim \frac{z_3(z_3-1)\Big((1
+ {\cal P}) h_{1,0} + {\cal P}
\rho_L\Big)+(1-z_3)\Big(h_{-1,0}
- {\cal P} \rho_L\Big)}{\sqrt{z_3\left(z_3-z_2\right)\left(z_4-z_3\right)\left(z_3-z_1\right)}}\frac{1}{\sqrt{\pi
n}}\frac{1}{z_3^n}
\end{equation}
For notational convenience, the latter expression can be rewritten as

\begin{equation}
h_{n,0}\sim K_+ \frac{\exp\Big(- n/\lambda_+\Big)}{n^{1/2}}
\end{equation}
where the decay amplitude is given by
\begin{equation}
\label{K+}
K_+=\frac{(z_3-1)\Big((1
+ {\cal P}) h_{1,0} + {\cal P}
\rho_L\Big)+(1/z_3-1)\Big(h_{-1,0}
- {\cal P} \rho_L\Big)}{\sqrt{8\pi}}  \left[\Big(\frac{1 + \sqrt{1 + {\cal P}}
 - \zeta}{\zeta}\Big)^2 - (1 + {\cal P})\right]^{-1/4},
\end{equation}
while the characteristic decay length obeys
\begin{equation}
\label{lambda+}
\lambda_+=1/\ln(z_3)
\end{equation}
Therefore, the density in front of the 
stationary moving tracer approaches the equilibrium value $\rho_L$
exponentially with the distance.

Note now that the decay 
amplitude $K_+$ and the characteristic length $\lambda_+$, 
Eqs.(\ref{K+}) and (\ref{lambda+}), depend on $\zeta$ and on $h_{\pm 1,0}$. 
It may be thus instructive to analyze the asymptotical forms 
of $K_+$ and $\lambda_+$, using the 
explicit results for  $h_{\pm 1,0} $ obtained in the previous
section. This analysis is presented in Appendix 2.

\subsection{Asymptotic forms of the density profiles at large distances 
past the stationary moving tracer.}

We first note that  one of the  roots of the generating function, 
namely $z_2$, gets equal to unity when both $f$ and $g$ are strictly 
equal to zero, which results in the exact cancellation of the multiplier $1 - z_3$
both in the nominator and the denominator in Eq.(\ref{h(z)})
. This shows that 
in the limit when exchanges with the reservoir are forbidden, 
qualitative changes in the singular behavior of the generating function at
the vicinity of $z_2$ will appear as compared to the case when $f,g >0$. 
Consequently, we have to consider separately 
the behavior in the case of non-conserved particles number, when exchanges with the
reservoir persist, and the case when both $f$ and $g$ are equal to
 zero while their ratio is kept fixed.

{\em A. Non-conserved particles number}.  
From Eq.(\ref{h(z)}) one finds that in the vicinity of $z = z_2$
the generating function $h(z)$ behaves as
\begin{equation}
h(z)=
\frac{z_2(z_2-1)\Big((1
+ {\cal P}) h_{1,0} + {\cal P}
\rho_L\Big)+
(1-z_2)\Big(h_{-1,0}
- {\cal P} \rho_L\Big)}{\sqrt{\left(z_2-z_1\right)\left(z_3-z_2\right)
\left(z_4-z_2\right)}}\frac{1}{\sqrt{z-z_2}}+{\mathcal O}(\sqrt{z-z_2}),
\end{equation}
which implies the following form of $h_{n,0}$ in the limit $n \to - \infty$,
\begin{equation}
\label{non}
h_{n,0}\sim K_-  \frac{\exp\Big(n/\lambda_-\Big)}{(-n)^{-1/2}},
\end{equation}
with
\begin{equation}
\label{K-}
K_-=\frac{(z_2-1)\Big((1
+ {\cal P}) h_{1,0} + {\cal P}
\rho_L\Big)+(1/z_2-1)\Big(h_{-1,0}
- {\cal P} \rho_L\Big)}{\sqrt{8\pi}}  \left[\Big(\frac{1 + \sqrt{1 + {\cal P}}
 - \zeta}{\zeta}\Big)^2 - (1 + {\cal P})\right]^{-1/4},
\end{equation}
and
\begin{equation}
\label{lambda-}
\lambda_- = -1/ln(z_2)
\end{equation}

{\it B. Conserved particles number.} Suppose now that both $f$ and $g$ are equal
 to zero, i.e. $\delta=0$, while their ratio is
fixed and given by $f/g = \rho_L/(1-\rho_L)$.
 As we have already remarked, this
 situation corresponds to the
usual model of a two-dimensional hard-core lattice gas without exchanges with a 
reservoir. Here, 
the generating function in the vicinity of its singular point $z = z_2 = 1$ obeys
\begin{equation}
h(z) = \left[\frac{2 {\cal P} \rho_L + (1 + {\cal P}) h_{1,0} - h_{-1,0}}{\sqrt{z_3-1}}-
h_{0,1} \sqrt{z_3-1}\right]\frac{1}{\sqrt{(1-z_1)(z_4-1)}}\sqrt{z-1}+{\mathcal O}((z-1)^3/2),
\end{equation}
which yields the following result for $h_{n,0}$ in the limit $n \to -\infty$,
\begin{equation} 
\label{a}
h_{n,0}\sim - \frac{K_-}{n^{3/2} }\left(1
+\frac{3}{8n}+{\mathcal O}\left(\frac{1}{n^2}\right)\right),
\end{equation}
where 
\begin{equation}
\label{am}
K_- = \frac{1}{4\sqrt{\pi}}\left(\frac{2 {\cal P} \rho_L + (1 + {\cal P}) h_{1,0} -
h_{-1,0}}{\sqrt{{\cal P}}}-h_{0,1}\sqrt{{\cal P}}\right)
\end{equation}
Remarkably enough, in this case the correlations between the tracer position
 and the particles distribution
vanish only {\it algebraically} slow with the distance! This implies, in turn,
 that in the conserved particles
number case, the mixing of the monolayer is not efficient and the medium "remembers"
 the passage of the tracer
for a long time which signifies strong memory effects. We note also that the algebraic
 decay of correlations 
in this model has been predicted earlier in Ref.\cite{deconinck}. However, the
 decay exponent has been
erroneously predicted to be equal to $1/2$, as opposed to the value
 $3/2$ given by Eq.(\ref{a}).  As well, the amplitude $K_-$ happens to have a different sign, compared to that in Ref.\cite{deconinck},
which invalidates the conclusion 
that the overall relaxation 
to the equilibrium value $\rho_L$ is a  non-monotonic function of the distance.

Asymptotic forms of $K_-$ and $\lambda_-$ in different limiting cases are
presented in Appendix 2.

\subsection{Integral characteristics of the density profiles}

First of all, we address the question whether the driven tracer, which induces
 an inhomogeneous density distribution
in the monolayer, 
shifts the equilibrium between adsorption and desorption, 
i.e. whether it changes effectively
the equilibrium density in the monolayer. The answer is trivially "no"
 in the case when the particles number is
explicitly conserved, but in the general case 
with arbitrary $f$ and $g$ this is not at all evident: desorption events are certainly 
favored in front of the tracer, while  the adsorption events are evidently suppressed by the excess density.
On the other hand, past the tracer desorption is diminished due to 
the particles depletion while adsorption
may proceed more readily due to the same reason.
 It is thus not at all clear $\it a \; priori$ whether these two effects can compensate
each other exactly, in view of the asymmetry of the density profiles.

For this purpose, we study first  the behavior of the integral
deviation $\Omega$ of the density from the equilibrium value $\rho_{L}$, i.e.
\begin{equation}
\Omega=\sum_{n=-\infty}^{+\infty}\sum_{m=-\infty}^{+\infty}h_{n,m}
\end{equation}
As a matter of fact, this integral characteristic can be computed straightforwardly from
 Eq.(\ref{expression fonction generatrice}). Setting $z$ and $w$ equal to unity, we find that
even in the case when $\delta \neq 0$, the integral deviation
$\Omega = 0$.  This means that the inhomogeneity
 of the density distribution in the monolayer created by the tracer
does not perturb the global balance between the adsorption and desorption events. 
 An analogous  result has been obtained for the
one dimensional problem in Ref.\cite{olivier}. 

Next we would like to check whether such a
 compensation holds on the axis of the tracer motion, i.e.
whether the integral deviation $\Omega_{X}$ 
along the $X$-axis is zero or not.  Setting $z = 1$ in
Eq.(\ref{h(z)}), we find then the following general
 result
\begin{equation}
\label{c}
\Omega_{X}= \sum_{n=-\infty}^{\infty} h_{n,0} = - \sqrt{\frac{\delta}{1+\delta}} \; h_{0,1},
\end{equation}
which implies that such an exact
 compensation is not realized
 on the $X$-axis only, except for the
case $\delta = 0$, which also
 does not seem to be a trivial
fact. Moreover, it follows from Eq.(\ref{c})
that $\Omega_{X}$ is always negative, provided that $ h_{0,1}$
 is positive definite, which means
apparently that on the
$X$-axis the equilibrium is shifted towards desorption.
 Note, however, that $\Omega_{X}$ always remains small and tends to zero in the limits 
$D_{0} \to 0$ and $D_{0} \to \infty$; hence,  $\Omega_{X}$  should be a bell-shaped function of the particles diffusivity.

Finally, we look at the total excess 
"weight" of the condensed region in front of the tracer particle, defined as
\begin{equation}
\Omega_{X+} = \sum_{n=1}^{\infty} h_{n,0}
\end{equation}
Using the expression for $h_{n,m}$ in Eq.(\ref{h en fonction de F}),
we find for $\Omega_{X+}$
the following asymptotical results:\\
(a) In the limit of small particles diffusivity and $\delta \neq 0$ (${\cal P} \gg 1$), 
\begin{equation}
\Omega_{X+} \sim \frac{\rho_{L}(1 - \rho_{L})}{\rho_{L} + (f +
g)\tau/\tau^*}.
\label{eq:10}
\end{equation}
This quantity is smaller than one, but  finite in the general case. Moreover, it appears
to be non-monotonic with respect to $\rho_L$. Lastly, it tends to zero
if $\tau^*$ tends to zero, i.e. when the exchanges with the vapour
phase proceed infinitely fast.\\ 
(b) In the limit of small particles diffusivity, $\delta = 0$ and ${\cal P} \ll 1$,
$\Omega_{X+} \sim 1$,
which result can be obtained from the previous one (\ref{eq:10}) by
taking the limit $\delta \to 0$, and recollecting that  $\rho_L\ll1$ in the limit
(b).\\
(c)In the limit of small particles diffusivity, $\delta = 0$ and ${\cal P} \gg 1$,
\begin{equation}
\Omega_{X+} \sim \frac{(1+\alpha )(1 - \rho_{L})}{\alpha \pi }
ln\Big(\frac{1}{1 - \rho_{L}}\Big).
\label{eq:12}
\end{equation}
Since here $\rho_L\sim1$, $\Omega_{X+}$ appears to be  small.\\
(d) In the limit of fast particles diffusion (${\cal P} \gg 1$),
$\Omega_{X+}$ approaches a constant value
\begin{equation}
\Omega_{X+} \sim \frac{(1+\alpha)\sigma^2 \rho_{L} ( 1-\rho_{L})}{\pi
 D_{0} \tau } ln\Big(\frac{D_{0}
 \tau}{\sigma^2 (1 -
\rho_{L})}\Big).
\label{eq:13}
\end{equation}
Note that  $\Omega_{X+}$ is non-monotonic with respect to
$\rho_L$, which generalizes the result announced in \cite{deconinck}
to the case  with non-conserved particles number. Note also that 
$\Omega_{X+}\to0$ as $D_0\to\infty$, which shows that the
condensed region disappears if the diffusion processes become
efficient enough to mix the adsorbed monolayer.

\section{Conclusions}

To conclude,  we have studied  analytically 
 dynamics of a driven probe molecule in a two-dimensional
adsorbed  monolayer composed of mobile, 
hard-core particles undergoing continuous exchanges 
with the vapour phase. Our analytical approach was based on
the master equation, describing the time
 evolution of the system, 
which allowed us to evaluate a system of coupled
dynamical equations 
for the tracer
particle velocity and a 
hierarchy of correlation functions. 
To solve these coupled equations, we have invoked an approximate closure scheme
based on the decomposition of the
third-order correlation functions into a product of pairwise correlations, which has
been  
first 
introduced in Ref.\cite{burlatsky} for  a related
 model of driven tracer dynamics in a one-dimensional lattice gas 
with conserved particles number. 
Within the framework of this approximation,
we have derived a system of coupled, discrete-space equations describing evolution 
of the density profiles, as seen from the  moving
probe, and its velocity $V_{tr}$. We have shown then that 
 $V_{tr}$ depends on the monolayer particles density in front of the tracer,
which is itself dependent on the
 magnitude of the velocity, 
as well as on the rate of the adsorption/desorption processes and the
 rate at which the particles can diffuse away of the tracer. 
As a consequence of such a non-linear coupling,    
in the general case, (i.e. for arbitrary adsorption/desorption 
rates and particles diffusion coefficient), 
$V_{tr}$ has been found only implicitly, 
as the solution of a certain non-linear 
equation relating its value to the system parameters. 
This equation 
simplifies considerably   
in the limit of small or large 
particles diffusivity, in which two cases
the tracer velocity is calculated explicitly. 
Further on, we have found that  the density profile  around the tracer becomes strongly
inhomogeneous: the local density of the monolayer
 particles in front of the tracer is higher than the 
average and approaches the average value as an exponential
 function of the distance from the tracer. The characteristic
length and the amplitude of the density relaxation function
are calculated explicitly.
On the other hand, past the tracer 
the local density is lower than the average; we show that depending on
 the condition whether the number of particles in the monolayer 
is explicitly conserved or not, the local density past the tracer
 may tend to the average value either as an exponential or  as an
 $\it algebraic$ function of the distance, revealing in the latter case
especially strong memory effects and strong 
correlations between the particle distribution in the
 monolayer and the tracer position.

\vspace{0.4in}

{\Large  \bf Acknowledgments.}

The authors wish to thank J.De Coninck, A.Lemarchand and A.A.Ovchinnikov
 for  helpful discussions. This work was supported in part by the 
French-German collaborative research
program PROCOPE.

\section{Appendix 1: Series representation of $F_{n,m}$,
Eq.~(\ref{calculdeF}), in the {\it {\bf limit (a)}}}

 We calculate to the first 
order in powers of $D_{0}\tau^*/\sigma^2$ the functions  $ F_{n,m}$ involved in
Eq.(\ref{speed}):

\begin{equation}
F_{0,0} =  1 + 
2 \frac{V^{(0)} \sigma }{\tau^*\Big(V^{(0)}  + \sigma (f + g)/\tau^{*}\Big)^2}\frac{D_0\tau^*}{\sigma^2}
+ {\mathcal O}\Big((\frac{D_{0} \tau^*}{\sigma^2})^2\Big),
\label{eq:1}
\end{equation}

\begin{equation}
F_{1,0} =F_{0,1}=  
\frac{ \sigma }{\tau^*\Big(V^{(0)} + \sigma (f + g)/\tau^{*}\Big)}\frac{D_0\tau^*}{\sigma^2}
+{\mathcal O}\Big((\frac{D_{0} \tau^*}{\sigma^2})^2\Big),
\label{eq:2}
\end{equation}

\begin{equation}
F_{-1,\pm1}=  
2\frac{ V^{(0)}\sigma }{\tau^*\Big(V^{(0)} + \sigma (f + g)/\tau^{*}\Big)^2}\frac{D_0\tau^*}{\sigma^2}
+{\mathcal O}\Big((\frac{D_{0} \tau^*}{\sigma^2})^2\Big),
\label{eq:3}
\end{equation}

\begin{equation}
F_{-1,0} = 
\frac{ V^{(0)} }{\Big(V^{(0)} + \sigma (f + g)/\tau^{*}\Big)}+
\frac{(f+g)\sigma^2\Big((f+g)(V^{(1)}+\sigma/\tau^*)+V^{(0)}(
V^{(1)}\tau^*/\sigma-2)\Big)}{\tau^{
*2}\Big(V^{(0)} + \sigma (f + g)/\tau^{*}\Big)^3}\frac{D_0\tau^*}{\sigma^2}
+{\mathcal O}\Big((\frac{D_{0} \tau^*}{\sigma^2})^2\Big),
\label{eq:4}
\end{equation}
and

\begin{eqnarray}
F_{-2,0} &=& 
\frac{ V^{(0)2} }{\Big(V^{(0)} + \sigma (f + g)/\tau^{*}\Big)^2}+\nonumber\\
&+&2
\frac{\sigma V^{(0)}\Big((f+g)^2(V^{(1)}\sigma/\tau^*+(\sigma/\tau^*)^2)+(f+g)V^{(0)}(V^{(1)}-2\sigma/\tau^*)-V^{(0)2}\Big)}{\tau^
*\Big(V^{(0)} + \sigma (f + g)/\tau^{*}\Big)^4}\frac{D_0\tau^*}{\sigma^2}+\nonumber\\
&+&{\mathcal O}\Big((\frac{D_{0} \tau^*}{\sigma^2})^2\Big),
\label{eq:5}
\end{eqnarray}

while

\begin{equation}
F_{2,0}=F_{1,\pm1}=F_{0,2}={\mathcal O}\Big((\frac{D_{0}
\tau^*}{\sigma^2})^2\Big)
\label{eq:6}
\end{equation}
   
\section{Appendix 2: Explicit results for the amplitudes and the decay 
lengths of the density profiles at large
distances }

\subsection{In front of the tracer}

We find the following explicit asymptotical forms for the decay length and the
amplitude:

(a) For small $D_{0}$, $\delta > 0$ and ${\cal P} \gg 1$,
\begin{equation}
\lambda_+ \sim   ln^{-1}\Big(\frac{\sigma^2 (f + g)(1+(f+g)\tau/\tau^*)}{D_{0} \tau^*(\rho+(f+g)\tau/\tau^*)} \Big),
\end{equation}
and 
\begin{equation}
K_+ \sim \frac{\rho_L (1 - \rho_L) (f + g)^{3/2}}{2 \sqrt{\pi}} \frac{(1 + 
(f + g) \tau/\tau^*)^{3/2}}{\Big(\rho_L + (f + g) \tau/\tau^*\Big)^{5/2}} \Big(\frac{\sigma^2}{D_0
\tau^*}\Big)^{3/2},
\end{equation}
which means that $\lambda_+$ is logarithmically small with $D_{0}$, 
while $K_+$ is large.  Note also that
$K_{+}$ can be a non-monotonous function of $\rho_{L}$ when $(f + g)
\tau/\tau^* \gg \rho_L$.\\
(b)  For small $D_{0}$, $\delta = 0$ and ${\cal P} \gg 1$,
\begin{equation}
\lambda_+ \sim   \frac{1}{ln{\cal P}} = ln^{-1}\Big(\frac{2 (1 -\rho_{L}}{\rho_{L}} \Big),
\end{equation}
and
\begin{equation}
K_+ \sim \frac{{\cal P}^{3/2}}{4 \sqrt{2} \pi},
\end{equation}
i.e., $\lambda_+$ is logarithmically small with ${\cal P}$, 
while $K_+$ is large.\\
(c)  For small $D_{0}$, $\delta = 0$ and ${\cal P} \ll 1$,
\begin{equation}
\lambda_+ \sim 1/{\cal P} =  \frac{\alpha \rho_{L}}{1 - \rho_{L}}; 
\end{equation}
and
\begin{equation}
K_+ \sim  \frac{1}{\pi} (1 + \alpha) \rho_L {\cal P}^{3/2} = \frac{1}{\pi} (1 + \alpha) 
\frac{(1 - \rho_{L})^{3/2}}{\alpha^{3/2} \rho_L^{1/2}}
\end{equation}
Hence, in this case $\lambda_{+}$ is large, 
since it is inversely proportional to the small parameter ${\cal P}$, but
$K_{+}$ is small.\\
(d) For $D_{0} \to \infty$ and $\delta > 0$,
\begin{equation}
\label{1}
\lambda_+ \sim \frac{1}{2 \sqrt{\delta}} = \sigma^{-1} \sqrt{\frac{D_{0} \tau^*}{(f + g)}}
\end{equation} 
and
\begin{equation}
\label{2}
K_+ \sim (1 + \alpha) \rho_{L} {\cal P} \Big(\delta/\pi^2\Big)^{1/4} = \frac{(1 + \alpha) \rho_L (1 -
\rho_L) \sigma^2}{\sqrt{\pi} D_0 \tau}  \Big(\frac{\sigma^2 (f + g)}{4 D_0 \tau^*}\Big)^{1/4},
\end{equation}
i.e., here, likewise to the case (c), $\lambda_+$ appears to be large 
and $K_+$ is small. On the other hand,  if we suppose that $\delta = 0$ (no adsorption/desorption)
and $D_0 \to \infty$, we find a bit different forms of $K_+$ and $\lambda_+$; namely,
\begin{equation}
\label{10}
\lambda_+ \sim \frac{1}{{\cal P}} = \frac{D_0 \tau}{(1 - \rho_L) \sigma^2}
\end{equation}
and
\begin{equation}
K_+ \sim \frac{(1+\alpha) \rho_{L} {\cal P}^{3/2}}{\sqrt{\pi}} = \frac{(1+\alpha) \rho_{L}}{\sqrt{\pi}}
\Big(\frac{\sigma^2 (1 - \rho_{L})}{D_0 \tau}\Big)^{3/2},
\end{equation}
i.e., similarly to the above considered case,  $\lambda_+$ is large and $K_+$ is small. Note also that, 
despite the fact that the tracer velocity in the cases
$\delta > 0$ and $\delta=0$ is given by the same expression in Eq.(\ref{vel}), the characteristic properties of the
density profile appear to be different when $\delta$ is zero or $\delta \neq 0$.

\subsection{Past the tracer}

{\em A. Non-conserved particles number}. 

Consider next the asymptotical behavior of $K_-$ and $\lambda_-$ in
 the limit of slow and fast particles
diffusion.\\
For small $D_{0}$, $\delta > 0$ and ${\cal P} \gg 1$, which corresponds to the limit (a) in our previous
notations,
\begin{equation}
\lambda_- \sim - ln^{-1}\Big(\frac{(1 - \rho_L)}{1 + (f + g) \tau/\tau^*}\Big),
\end{equation}
and
\begin{equation}
K_- \sim  - \frac{\rho_L }{2 \sqrt{\pi}} \Big(\frac{(f+g) (1 + (f+g)\tau/\tau^*)}{\rho_L
 + (f+g)\tau/\tau^*} (\frac{\sigma^2}{D_0 \tau^*})\Big)^{1/2},
\end{equation}
i.e. $\lambda_-$ tends to a constant value 
when $D_0 \to 0$, which means that here $\lambda_- > \lambda_+$, while
$|K_-| \propto  D_{0}^{-1/2}$  diverges when $D_{0} \to 0$. Note, however, that we evidently
have $|K_-| ~ \ll K_+$.  This means that in the limit (a) 
the density profiles around the tracer are strongly asymmetric;
in front of the tracer we have a condensed, "traffic"-jam-like  region, characterized by a high amplitude
but of a relatively short spatial extent, while past the tracer there is a depleted region which extends on much
longer scales but  has a considerably smaller amplitude.\\
Next, for high particles diffusivity and $\delta > 0$, which corresponds to the limit (d), one encounters an
exactly opposite situation. Here the values of the characteristic
lengths and the amplitudes of the condensed and the
depleted regions almost coincide, i.e. 
  $K_- \approx - K_+$ and $\lambda_-
\approx \lambda_+$, where $K_+$ and $\lambda_+$ are given by Eqs.(\ref{2}) and
(\ref{1}).

{\em B. Conserved particles number.}

Consider next the asymptotic forms of the decay amplitudes in Eq.(\ref{am}). Using the results of the
previous section, we find then the following asymptotical results:\\
In the limit of small $D_{0}$, $\delta = 0$ and ${\cal P} \gg 1$, 
 which corresponds to the limit (b) of Section
5, 
\begin{equation}
K_- \sim  \frac{{\cal P}^{1/2}}{4 \sqrt{\pi}} = \Big(\frac{(1 - \rho_L)}{8 \pi \rho_{L}}\Big)^{1/2},
\end{equation}
which means that the amplitude is large in this limit, since $\rho_{L} \ll 1$.\\
In the limit of small $D_{0}$, $\delta = 0$ and ${\cal P} \ll 1$, (limit (c)), 
\begin{equation}
K_- \sim  \frac{(1 + \alpha)}{2}  \Big(\frac{\rho_{L}(1 - \rho_L)}{\pi \alpha}\Big)^{1/2}
\end{equation}
Since this limit can be only realized at sufficiently high particles densities, $\rho_L \sim 1$,
we have then that in this case the amplitude should be small.\\
Lastly, in the limit $D_{0} \to \infty$ (limit (d)) we find 
\begin{equation}
K_- \sim \frac{(1+\alpha) \rho_L}{2} \Big(\frac{(1-\rho_l) \sigma^2}{\pi D_0 \tau}\Big)^{1/2},
\end{equation}
i.e. which signifies that in this case the decay amplitude is small.

\newpage

\begin{center}
\resizebox{!}{!}{\includegraphics[scale=0.6]{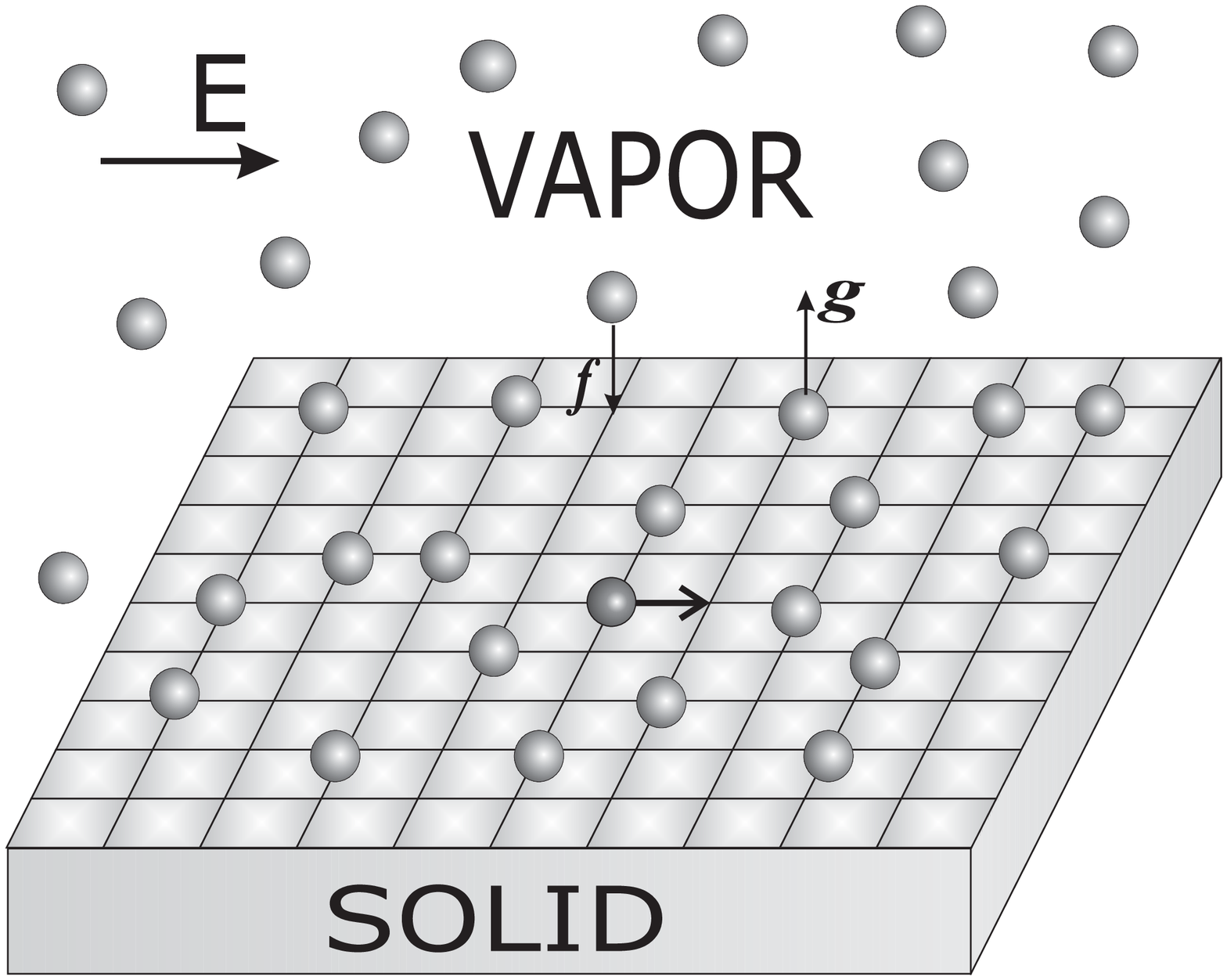}}
\\
\vspace{2in}
Fig.1.  Two-dimensional lattice of adsorption sites 
partially occupied by identical, mobile hard-core particles 
(grey spheres) 
undergoing continuous exchanges with the reservoir - the vapor phase.
Particles desorption and adsorption probabilities are denoted by $g$
and $f$, respectively. 
The dark grey sphere with an arrow 
denotes the tracer particle, whose motion is
completely directed by external field $E$.
\newpage

\resizebox{!}{!}{\includegraphics{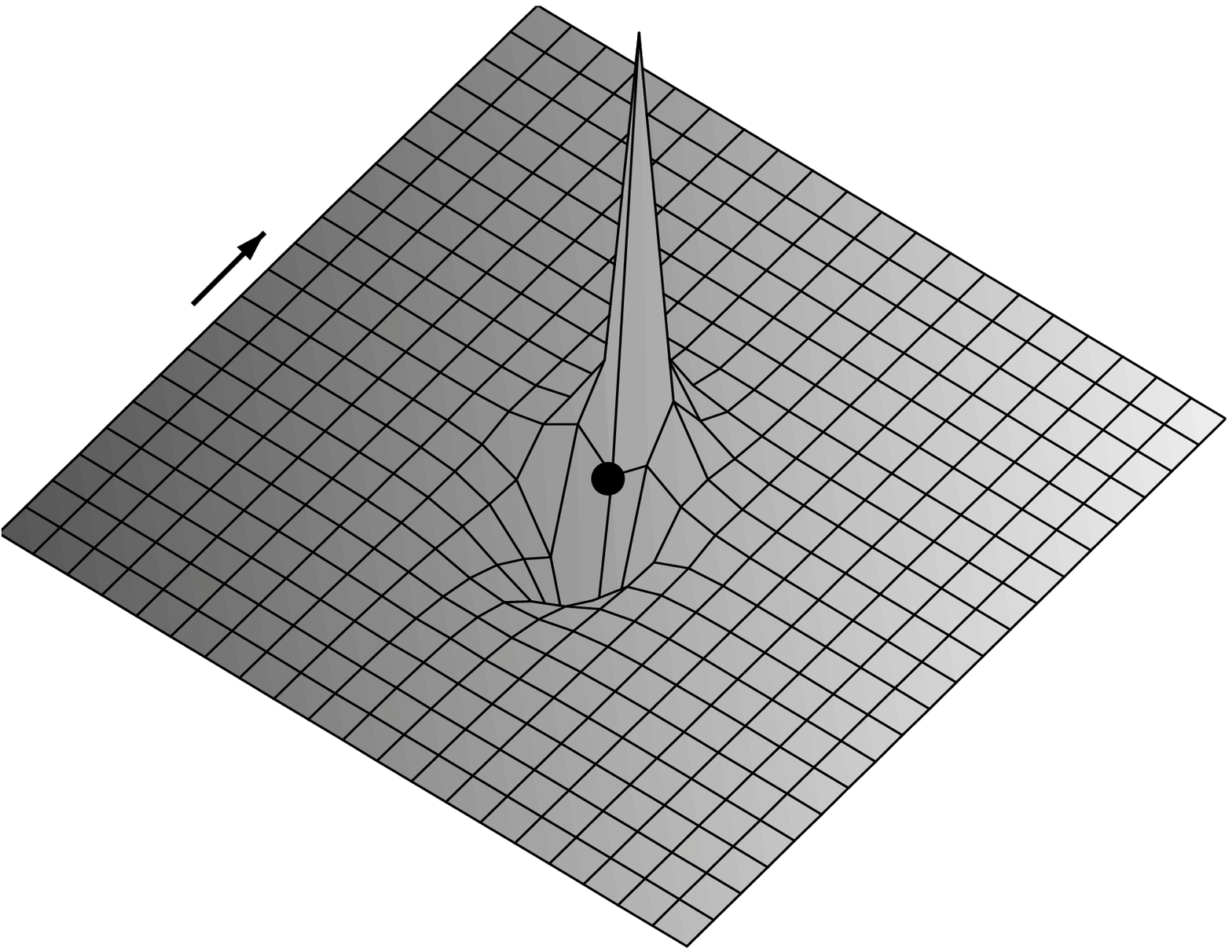}}
\\
\vspace{2in}
Fig.2a.  Sketch of the density profiles at finite distances from the 
tracer. The black circle
denotes the tracer particle and the arrow represents the direction of
the external force exerted on the tracer. {\it Limit (a)}: 
$D_0\to0,\delta\gg1$, and ${\cal P}\gg1$.

\newpage

\resizebox{!}{!}{\includegraphics{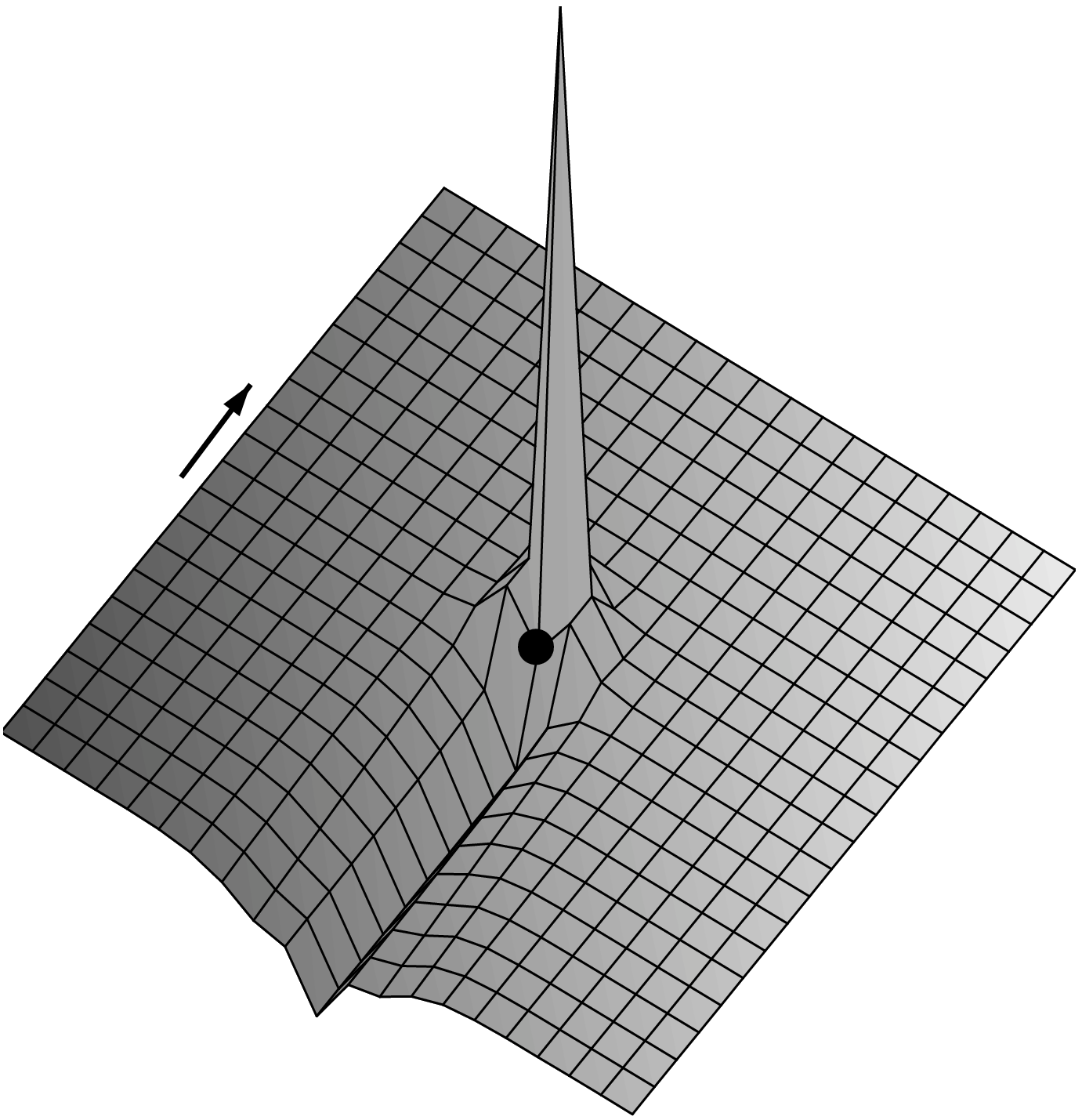}}
\\
\vspace{2in}
Fig.2b. Sketch of the density profiles at finite distances from the 
tracer. {\it Limit (b)}: 
$D_0\to0,\delta=0$, and ${\cal P}\gg1$.
\newpage

\resizebox{!}{!}{\includegraphics{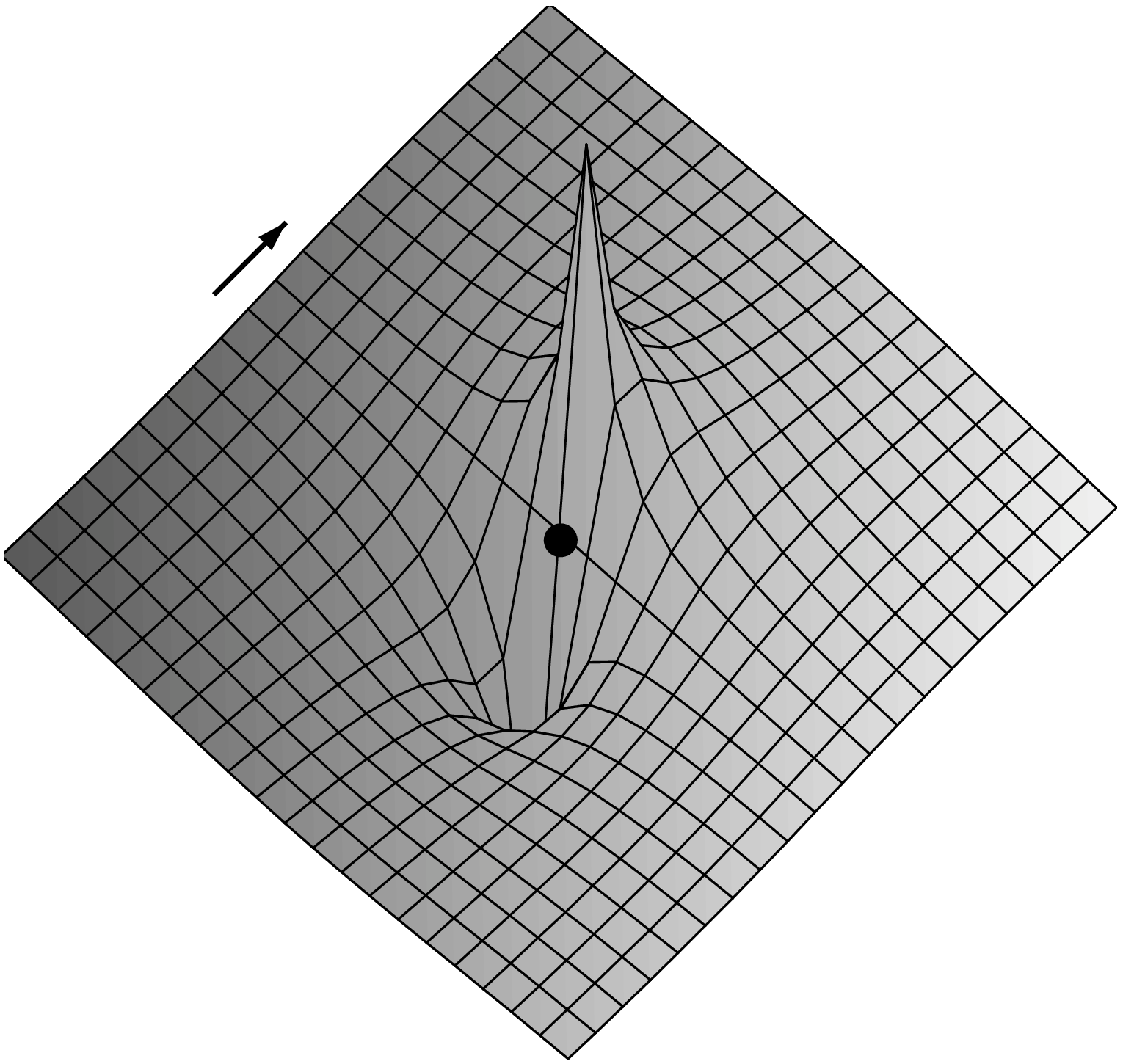}}
\\
\vspace{2in}
Fig.2c. Sketch of the density profiles at finite distances from the 
tracer  {\it Limit (c)}:
$D_0\to0,\delta=0,$ and ${\cal P}\ll1$.

\end{center}

\end{document}